\documentclass[twocolumn,twocolappendix]{aastex631}
\shorttitle{Tidal CHE and HD\,5980}
\shortauthors{Sharpe et al.}
\graphicspath{{./}{figures/}}

\usepackage{xspace} %this package handles when to use a whitespace in math symbols
\usepackage{natbib} %this package can be used for creating an automated bibliography. 
\usepackage[capitalize]{cleveref}
\usepackage{url}
\usepackage{hyperref}

\newcommand{\Rsun}{\ensuremath{\,\mathrm{R_\odot}}\xspace} % Solar radius symbol
\newcommand{\Msun}{\ensuremath{\,\mathrm{M_\odot}}\xspace} % Solar mass symbol
\newcommand{\Myr}{\ensuremath{\,\mathrm{Myr}}\xspace} % symbol for Mega year
\newcommand{\days}{\ensuremath{\,\mathrm{d}}\xspace} % symbol for day

\newcommand{\MESA}{{\tt MESA}\xspace}

\begin{document}

\title{Investigating the Chemically Homogeneous Evolution Channel and its Role in \\ the Formation of the Enigmatic Binary Black Hole Progenitor Candidate HD\,5980
}

\author[0000-0001-8225-8969]{K. Sharpe}
\affiliation{Department of Astronomy, University of California, Berkeley, 501 Campbell Hall \#3411, Berkeley, CA 94720, USA}
\affiliation{Center for Astrophysics | Harvard \& Smithsonian, 60 Garden Street, Cambridge, MA 02138, USA}
\affiliation{Max-Planck-Institut f\"ur Astrophysik, Karl-Schwarzschild-Straße 1, 85741 Garching, Germany} 

\author[0000-0001-5484-4987]{L.A.C. van Son}
\affiliation{Center for Computational Astrophysics, Flatiron Institute, New York, NY 10010, USA}
\affiliation{Center for Astrophysics | Harvard \& Smithsonian, 60 Garden Street, Cambridge, MA 02138, USA}
\affiliation{Anton Pannekoek Institute for Astronomy and GRAPPA, University of Amsterdam, NL-1090 GE Amsterdam, The Netherlands} 
% \affiliation{Max-Planck-Institut f\"ur Astrophysik, Karl-Schwarzschild-Straße 1, 85741 Garching, Germany}

\author[0000-0001-9336-2825]{S. E. de Mink}
\affiliation{Max-Planck-Institut f\"ur Astrophysik, Karl-Schwarzschild-Straße 1, 85741 Garching, Germany}
\affiliation{Anton Pannekoek Institute for Astronomy and GRAPPA, University of Amsterdam, NL-1090 GE Amsterdam, The Netherlands} 

\author[0000-0003-3441-7624]{R. Farmer}
\affiliation{Max-Planck-Institut f\"ur Astrophysik, Karl-Schwarzschild-Straße 1, 85741 Garching, Germany}

\author[0000-0002-0338-8181]{P. Marchant}
\affiliation{Institute of Astronomy, KU Leuven, Celestijnenlaan 200D, 3001 Leuven, Belgium}

\author{G. Koenigsberger}
\affiliation{Instituto de Ciencias Físicas, Universidad Nacional Autónoma de México, Ave. Universidad s/n, Chamilpa 62210,
Cuernavaca, Mexico}

\begin{abstract}
Chemically homogeneous evolution (CHE) is a promising channel for forming massive binary black holes. The enigmatic, massive Wolf-Rayet (WR) binary HD\,5980 A\&B has been proposed to have formed through this channel. We investigate this claim by comparing its observed parameters with CHE models.  
Using \texttt{MESA}, we simulate grids of close massive binaries then use a Bayesian approach to compare them with the stars’ observed orbital period, masses, luminosities, and hydrogen surface abundances.
The most probable models, given the observational data, have initial periods $\sim$3 days, widening to the present-day $\sim$20 day orbit as a result of mass loss — correspondingly, they have very high initial stellar masses ($\gtrsim$150 \Msun). We explore variations in stellar wind-mass loss and internal mixing efficiency, and find that models assuming enhanced mass-loss are greatly favored to explain HD\,5980, while enhanced mixing is only slightly favoured over our fiducial assumptions. Our most probable models slightly underpredict the hydrogen surface abundances. 
Regardless of its prior history, this system is a likely binary black hole progenitor. We model its further evolution under our fiducial and enhanced wind assumptions, finding that both stars produce black holes with masses $\sim19-37\Msun$. The projected final orbit is too wide to merge within a Hubble time through gravitational waves alone. However, the system is thought to be part of a 2+2 hierarchical multiple. We speculate that secular effects with the (possible) third and fourth companions may drive the system to promptly become a gravitational-wave source.

\end{abstract}


\section{Introduction} \label{sec:intro}

Detections of gravitational waves resulting from binary black holes and neutron star mergers have begun to reveal the properties of the population of double compact objects \citep{GWTC-3_popPaper}. How these double compact objects form and how they end up in orbits tight enough to induce a gravitational wave inspiral are still unknown. Various promising scenarios have been proposed \citep[see reviews by][]{Mapelli2020_review, MandelBroekgaarden2022}, and it seems likely that several of them contribute to the population of double compact objects \citep[e.g.,][]{Wong+2021,Zevin+2021,Bouffanais+2021,StevensonClarke2022,Godfrey+2023}.
To advance our understanding and constrain their relative contributions, it is necessary to investigate the progenitors and side products of these scenarios, which may be detectable with conventional electromagnetic (EM) observations.  In this work, we investigate the constraints derived from EM observations for a very massive binary system, which has been suggested to be the result of chemically homogeneous evolution (CHE). The CHE channel is one of the promising contenders for the formation of massive binary black holes \citep{de-Mink+2009a, Mandel+2016, de-Mink+2016, duBuisson+2020, Riley+2021}.

CHE is a hypothesized mode of evolution where stars experience enhanced mixing. This allows hydrogen-rich material from the outer layers of the star to reach the central regions, where it is used to power core hydrogen burning. At the same time, helium produced in the central regions is mixed throughout the star, including the outer layers. The result is a star with a chemically homogeneous composition. Such stars stay compact and result in very massive helium stars. Eventually, after completing the subsequent burning stages, the star will most likely collapse and form a black hole \citep[e.g.][]{Maeder+1987, Yoon+2006}.  

 \citet{de-Mink+2009a} proposed that CHE may occur in very close binary systems, where enhanced mixing may be expected as a result of tidal deformation and tidal spin up \citep[cf.][]{Song+2013, Hastings+2020}. As the stars remain very compact, this type of evolution can prevent the two stars from merging and produce two very massive helium stars. This makes the CHE pathway of interest as a formation pathway for gravitational wave sources, as initially suggested independently by \citet{Mandel+2016} and \citet{Marchant+16}. 
 Since then, several studies have further explored this pathway \citep{de-Mink+2016, Marchant+2017, duBuisson+2020, Riley+2021, Ghodla+2023, Dorozsmai+2023} 

Chemically homogeneous evolution has also been considered in the case of very rapidly rotating single stars, where mixing processes resulting from rotation may be responsible for keeping the stars homogeneous \citep[][]{Maeder+1987, Yoon+2005, Yoon+2006, Brott+2011}. In wider binaries, CHE may also occur for the mass-gaining star that is spun up as a result of accretion \citep[e.g.,][]{Cantiello+2007,de-Mink+2013,Ghodla+2023}. 
In our work, we will refer to this as \emph{``birth-spin CHE''} and \emph{``accretion-induced CHE''} respectively, to distinguish them from \emph{``tidal CHE''} as described in the previous paragraph. 

In this work, we study the evolution of a very massive binary system that is arguably the most promising candidate for CHE: the massive WR binary HD\,5980 \citep[e.g.,][cf. a review of further candidates for CHE in Appendix \ref{app: further CHE}]{Koenigsberger+2014,Shenar+2016}. Our aims are (1) to investigate the claim made by \citet{Marchant+16} that this system can be explained as a system that evolved through tidal CHE,  (2) to explore the effect of variations in the uncertain mass loss and internal mixing on the channel, and their effect on our ability to match the observed properties of the system, (3) to infer its initial parameters and current evolutionary state under the assumption that it formed through tidal CHE, and (4) to speculate on its future evolution and final fate as a possible binary black hole system.  

Our paper is organized as follows. 
In Section \ref{sec:obs HD5980}, we provide an extensive description of HD\,5980 and summarize earlier work that considered a chemically homogeneous origin for this system.  In Section \ref{sec:method}, we describe the stellar evolutionary code and choices for the physical assumptions. In Section \ref{sec: CHE window}, we present our model grids then discuss the window for chemically homogeneous evolution and how it depends on our assumptions for mixing and mass loss. In Section \ref{sec: best fit for HD5980}, we discuss our Bayesian fitting procedure for comparing HD\,5980 with our models, and we present the posteriors for the physical parameters. In Section \ref{sec:future evol}, we speculate on the further evolution of the system. We argue that it is a likely progenitor of a binary black hole system, and we discuss the possible role of the third and fourth companion. In Section \ref{sec:limitations}, we provide a critical evaluation of the limitations of our models and discuss alternative evolutionary explanations. In Section \ref{sec:conclusion}, we provide a brief summary and conclusion.

\section{The Case of HD~5980 \label{sec:obs HD5980} }

\subsection{Observational constraints}

HD\,5980 is a remarkable multiple system in the SMC, which contains a massive WR binary system \citep[with component masses of about 60\Msun,][]{Koenigsberger+2014,Shenar+2016}. Table \ref{tab:params} provides an overview of the observed properties. 
Despite having been observed spectroscopically since the 1960s \citep{Feast+1960}, its evolutionary origin remains a puzzle to this day. 
The system contains a binary pair in a 19.3\,day orbit and a third light source which itself is in a $\sim97$\,day orbit around an undetermined companion \citep{Koenigsberger+2014}.
We will refer to the inner WR binary as HD\,5980 A and B and the third (light source) and fourth (unseen) objects as HD\,5980 C and D, respectively. We note that it has not been confirmed that the two binaries are gravitationally bound to each other, nor is this orbit characterized, but the literature classifies them under these naming conventions.

The inner binary, HD\,5980 A\&B, has an eccentric orbit, $e=0.27$, and a 19.3 day period \citep{Koenigsberger+2014}. 
Both stars exhibit high surface He-abundance (see Table \ref{tab:params}).
The inferred post-eruption effective temperature of star A lies in the range $40-45$\,kK \citep{Georgiev+2011, Shenar+2016, Hillier+2019}. 
However, given the strong mass-loss rates of HD\,5980, the stellar surface is effectively embedded within an extended, optically thick wind region, causing the observed value of $T_\mathrm{eff}$ to be significantly lower than the surface temperature comparable to those predicted by stellar structure models. For example, \texttt{MESA} models typically define the photosphere and $T_\mathrm{eff}$ at $\tau=2/3$, without modelling any optically thick winds. Because of this, we do not constrain our study based upon $T_\mathrm{eff}$ (see also the discussion in Section \ref{sec:limitations}).

HD\,5980 A has been observed to undergo eruptive outburst events, with two events observed in 1993 and 1994, each lasting less than a year \citep{Bateson+1994,BarbaNiemela1995,Barba+1995,Koenigsberger+1995, Sterken+1997}, and one inferred to have occurred in the 1960s \citep{Koenigsberger+2010}.
\cite{Georgiev+2011} note that these bursts could be driven by the hot and cool iron opacity bumps \citep[the bi-stability bump,][]{Lamers+1995,Vink+1999}. 
\cite{Koenigsberger+2014} further argue that these luminous blue variable (LBV)-type outbursts might be related to a super-Eddington layer that turns unstable \citep{Humphreys+1994,Stothers+1996}. Alternatively, \cite{Foellmi+2008} speculate that this eruptive behavior could be linked to gravitational perturbations from the periastron passage between the A$+$B and C$+$D binaries. The cause of the outburst and the system's unusual properties are not well understood, making HD\,5980 a focal point of numerous studies \citep[see][]{Breysacher+1980, Koenigsberger2004, Georgiev+2011, Shenar+2016, Hillier+2019, Naze+2018, Kolaczek-Szymanski+2021}.
The outburst also affected the observed properties, like the mass-loss rate and temperature of HD\,5980 \citep[e.g.,][]{Koenigsberger+1995, Georgiev+2011}. 
For the purposes of this paper, we consider only the quiescent, non-outburst parameters for star A.

\begin{table*}[ht]
\centering
\begin{tabular}{llll}
\hline 
      HD\,5980                   & \multicolumn{2}{c}{A\&B}               & \multicolumn{1}{c}{C\&D}     \\ \hline \hline
$P_{\rm orb}$ [days]             & \multicolumn{2}{c}{$19.2656\pm0.0009^a$} & \multicolumn{1}{c}{$96.56\pm0.01^{a}$} \\
eccentricity                     & \multicolumn{2}{c}{$0.27\pm0.02^{a}$}     & \multicolumn{1}{c}{$0.815\pm0.020^a$} \\ 
Inclination [$^\circ$]           & \multicolumn{2}{c}{$86\pm1^b$}           & \multicolumn{1}{c}{-}               \\
Age [Myr]                        & \multicolumn{3}{c}{$\approx 3\pm1^{\rm g}$}   \\ \hline \hline 

    Stellar properties           & \multicolumn{1}{c}{A}    & \multicolumn{1}{c}{B}      & \multicolumn{1}{c}{C}     \\ \hline \hline
Spectral type                    & WN6h              & WN6-7             & O \\     
$\log_{10}$ L                   & $6.35\pm 0.10^c$  & $6.25\pm 0.15^{c}$ & $5.85\pm 0.10^c$\\ 
$M_\mathrm{orb}$ [\Msun]         & $61\pm 10^{a,c}$ & $66\pm 10^{a,c}$   & \multicolumn{1}{l}{-} \\
$M_\mathrm{g}$ [\Msun]         & - & -   & \multicolumn{1}{l}{$34^{+64~c}_{-22}$} \\
X$_\mathrm{H}$                   & $0.25\pm 0.05^c$ & $0.25\pm 0.20^c$   & \multicolumn{1}{l}{-} \\
$X_\mathrm{He}$                              & $0.80^{d,e}$     &  $\sim X_{\mathrm{He}_A}$         & - \\
$v_{\rm surf}$ [$\rm km s^{-1}$] & $250^{f}$        & $75^f$         & \multicolumn{1}{l}{-}\\
$T_{*_{obs}}~\mathrm{[kK]} $    & $45_{-5}^{+5~c}$   & $45_{-7}^{+10~c}$     & $34_{-3}^{+3~c}$ \\
$T_{s}~\mathrm{[kK]} $    & $60^d$   & -     & $27^d$ \\
$\log_{10} \dot{M}$              & $-4.5 \pm 0.1^c$  & $-4.5 \pm 0.3^{c}$    & $-5.9^{c}$\\
\end{tabular}
\caption{Observed Parameters of HD\,5980, values obtained by a) \cite{Koenigsberger+2014}, b) \cite{Perrier+2009},  c) \cite{Shenar+2016}, d) \cite{Georgiev+2011}, e) \cite{Koenigsberger+1998}, and f) \cite{GeorgievKoenigsberger2004}. g) Estimate of age of NGC 346 from \cite{Mokiem+2006}, \cite{Sabbi+2007}, and citations therein.}
\label{tab:params}
\end{table*}

\subsection{A Chemically Homogeneous Origin}
HD\,5980 was first suggested to be evolving chemically homogeneously by \cite{Koenigsberger+2014}.
They argue that the progenitor stars required to explain the massive components of HD\,5980 \citep[about $120\Msun$ at the zero-age main sequence, ZAMS,][]{Koenigsberger2004} would not fit within its current Roche-lobe radius (of about $57\Rsun$). They further argue that HD\,5980 is difficult to explain with stable binary mass transfer given the large He surface abundance and rather evolved appearance of \textit{both} stars. This leaves CHE as one of the most viable solutions. 
They use the Binary Evolution Code \citep[][]{Heger+2000, Petrovic+2005a, Yoon+2006} to evolve a single system with $P_{\rm init} = 12\days$, ZAMS masses of 90\Msun and 80\Msun, and initial rotational velocities of 500\,km\,s$^{-1}$, and find that this birth-spin CHE can adequately explain many of the observed parameters of star A. 

\cite{Schootemeijer+Langer2018} reinforce the conclusions from \cite{Koenigsberger+2014}.
They compute two grids of stellar models at SMC metallicity using \MESA, one with low initial rotation velocities and one where the stars rotate rapidly at birth and hence evolve in part or completely chemically homogeneously. 
They conclude that HD\,5980 cannot be explained well by their models which experience stable mass transfer. Instead, it is best explained by models with high initial rotational velocity (about 520-540 $\rm km s^{-1}$) that evolve chemically homogeneously (birth-spin CHE). 
They find corresponding initial component masses of $70-80\Msun$, depending on if the components of HD\,5980 are currently core-hydrogen or core-helium burning. 

Lastly, \cite{Marchant+16} present an extensive model grid of near- and overcontact binary systems that lead to tidal CHE. Based on this grid, they suggest that HD\,5980 may be explained by tidal CHE, beginning as a close contact system and widening to its present-day state due to stellar wind mass-loss. An in-depth analysis of this scenario for HD\,5980 was outside the scope of their work. In this work we consider the last scenario: tidal CHE as the evolutionary explanation for HD\,5980.

\section{Method}
\label{sec:method}

We use the open source Modules for Experiments in Stellar Astrophysics, \texttt{MESA} \citep[][]{Paxton+2010,Paxton+2011,Paxton+2013,Paxton+2015,Paxton+2018,Paxton+19}, version 12778 to model the evolution of massive, initially closely orbiting binary star systems that evolve chemically homogeneously, closely following the approach of \cite{Marchant+16}. Below we describe the physical choices and set up of our grid.

\subsection{Physical Assumptions}

We adopt an initial metallicity of $Z = Z_\odot/4 = 0.00425$, using $Z_\odot = 0.017$ and solar-scaled abundances as in \cite{Grevesse+1996}. 
We use the \texttt{basic} nuclear network to follow H- and He-burning, which is sufficient for our purposes.
For convection, we use the Ledoux criterion and the standard mixing-length theory \citep{Bohm-Vitense1958} with a mixing length parameter of $\alpha = 1.5$. We model semiconvection as in \citet{Langer+1983}, and set the efficiency parameter $\alpha_{\rm sc} = 1.0$. Core overshooting during core hydrogen burning is incorporated as in \cite{Brott+2011}.

For the effect of tides on the orbital evolution, we adopt \cite{Hut1981a} for stars with a radiative envelope \citep[see][for details]{Paxton+2015}. We initiate our models such that the orbit is circular and the spins of the stars are synchronized to the orbit at ZAMS. This is a reasonable assumption because the synchronization timescale is only a very small fraction of the main-sequence lifetime \citep[see e.g., ][]{de-Mink+2009}, even when assuming less efficient tides \citep[such as e.g.,][]{Yoon+2010}.

Where applicable, we compute the contact binary phase following the approach of \citet[][]{Marchant+16}, including the treatment of L2 overflow.
We ignore possible heat exchange during over contact phases, which is appropriate when modelling equal mass twin stars \citep[][]{Fabry+2022, Fabry+2023}.
For the effect of magnetic fields on the transportation of angular momentum, we assume the  Spruit-Tayler dynamo \citep{Spruit2002}, implemented as described in \cite{Petrovic+2005a}.
The effect of the centrifugal acceleration on the stellar structure equations is accounted for following \cite{Kippenhahn+1970} and \cite{Endal+1976} as described in \cite{Paxton+19}.

\paragraph{Chemical mixing}
We account for mixing of chemical elements and the transport of angular momentum by rotational instabilities. Among the relevant processes, Eddington-Sweet circulation is the dominant process, which we treat as a diffusive process. We also allow for secular \citep{Endal+1978b} and dynamical \citep{Zahn1974a} shear instabilities and the Goldreich-Schubert-Fricke instability \citep{Goldreich+1967, Fricke1968}, using $f_c = 1/30$ following \cite{Endal+1976} and \cite{Pinsonneault+1989} as described in \cite{Heger+2000a}.
To account for the effects of tidal deformation on mixing in very close binary systems, we follow \citet{Hastings+2020} and increase the diffusion coefficient for Eddington-Sweet to $\rm D_{\rm ES} = 2.1$ for our fiducial model grid. 

The efficiency of chemical mixing is generally uncertain. We therefore explore two variations of $\rm D_{\rm ES}$: creating one model grid to represent decreased mixing, adopting $\mathrm{D}_\mathrm{ES}=1.0$ (labeled $\sim$$\times$1/2 mixing), and one model grid that represents increased mixing, where we set $\mathrm{D}_\mathrm{ES}=4.0$ ($\sim$$\times$2 mixing).

\paragraph{Stellar winds}
For stellar winds, we follow the implementation in \cite{Brott+2011} of \cite{Yoon+2006}. For hydrogen-rich stars with surface helium abundance $\rm{Y}_{\rm s} < 0.4$, the recipe is taken from \cite{Vink+2001}, while for hydrogen-poor stars with $\rm{Y}_{\rm s} > 0.7$, the prescription from \cite{Hamann+1995}, divided by a factor of ten, is used. For surface helium abundances between 0.4 and 0.7 ($0.4 < \rm{Y}_{\rm s} < 0.7$), the wind mass-loss rate is interpolated between the calculated values for \cite{Vink+2001} and \cite{Hamann+1995}. These recipes account for the reduction of the mass-loss rate at the metallicity of the SMC. 
Winds are enhanced due to rotation as in \cite{Heger+2000a}. 

Stellar wind mass-loss remains one of the main uncertainties in massive star evolution, and, in particular, the winds of very massive stars (with masses above $\sim100\Msun$) are poorly constrained by observations. 

Several works that focus on the optically thin winds of OB stars \citep[e.g.,][]{2007A&A...468..649L,2017A&A...606A..31K,2019A&A...632A.126S} report much lower mass-loss rates than those of \cite{Vink+2000}. 
\cite{Bestenlehner20} finds that the prescriptions from \cite{Vink+2000} under predict by a factor of a few the observed rates from WN5h stars, while heavily overpredicting the wind mass-loss rates from results from O8V stars.
This is also in line with results from \cite{Grafener+2021} and \cite{Sabhahit+2022}, who suggest that the wind mass-loss rates of very massive H-rich stars are higher than those predicted by \cite{Vink+2001}.
Additionally, for H-poor stars, \cite{SanderVink2020} and \cite{Sander+2020} argue that the prescription from \cite{Hamann+1995} will over-predict the wind mass-loss rate for lower mass stars, but might under-predict the rate for the highest mass stars.

In order to explore the effects of uncertainties in the wind mass-loss, we consider two variations with respect to our fiducial input physics assumption: a decreased winds variation, where we reduce \textit{all} wind mass-loss rates by a factor three (labeled $\times$1/3 Winds), and similarly an increased winds variation where we increase \textit{all} wind mass-loss rates by a factor three ($\times$3 Winds). 

We account for the effect of wind mass-loss on the orbit as in \citet{Paxton+2015}.
For the shortest period stages of evolution, this potentially underestimates angular momentum losses due to tidal coupling between the wind and the binary system \citep{MacLeod+2020}. However, due to the uncertainties on the strength of this effect, we ignore it in our evolution models. Additionally, as discussed in Section \ref{sec:intro}, HD\,5980 has been observed to experience eruptive mass-loss episodes in the past. Lacking a physical model for these outbursts, we assume that all mass is lost through steady winds (see also the discussion in Section \ref{sec:limitations}).

\subsection{Model Grid Setup} \label{sec:model_grids}
We evolve binary star models starting from the zero-age main sequence (ZAMS) for several two-dimensional model grids in the initial period and initial component mass. For each grid, we assume that both stars in the system have equal initial masses (q=1). The range of initial periods is 1.4 days to 5.0 days, with linear steps of 0.2 days. Initial component masses range from 1.9 to 2.45 in $\log(\mathrm{M}/\mathrm{M_\odot})$ ($\approx 79\Msun$ to $281\Msun$), with steps of 0.025 in log-space. The five computed model grids differ from each other by the value of $\rm D_\mathrm{ES}$ and the wind mass-loss rate, described in the previous section.

We allow models to run through core helium burning. However, due to difficulties with modeling such high-mass systems, most models terminated just before or during core helium ignition. Thus, in our analysis, we cut the models' evolution at the beginning of core He-burning (which we define as the time when the center helium mass fraction is decreasing and falls below 0.994). We do not follow the evolution of models that are not evolving chemically homogeneously. That is, we stop evolving systems when the central and surface helium mass fraction differs by more than 0.2 \citep[as in][]{Marchant+16}.

\section{The Window for Chemically Homogeneous Evolution 
\label{sec: CHE window}}

\begin{figure*}
\centering
\includegraphics[width=0.95\linewidth]{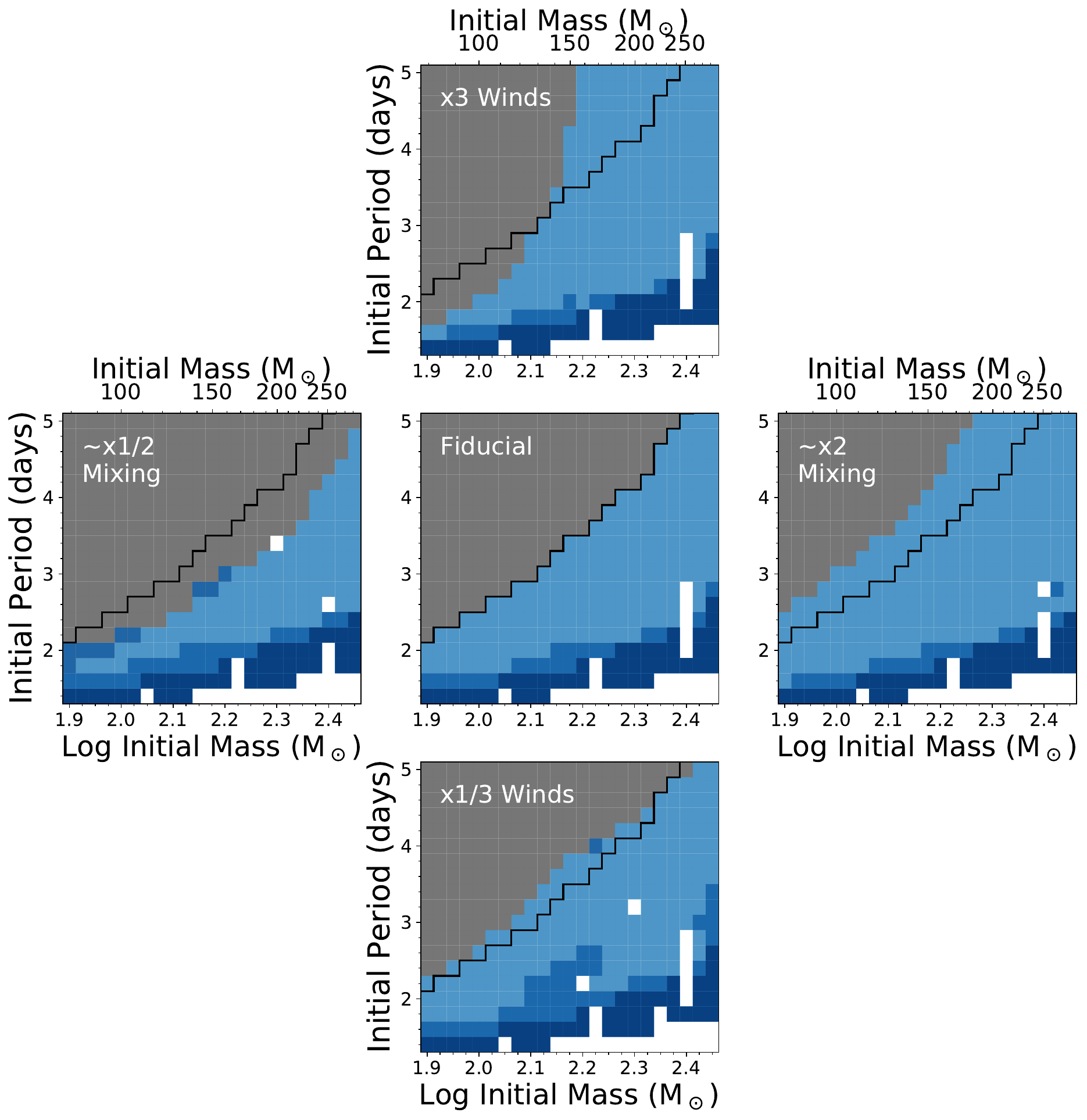}
\caption{The window for chemically homogeneous evolution in initial periods and masses, every coloured square represents one \texttt{MESA} model. Blue coloring corresponds to models that evolve chemically homogeneously. Shades of blue indicate when the models experienced RLOF; dark blue indicates RLOF at ZAMS, which occurs through the L1 point (L1OF), medium blue shows models which experience L1OF post-ZAMS, while the lightest blue shows models which do not experience any RLOF during the model run. We find that our models do not experience L2OF.
Models that do not evolve chemically homogeneously are colored grey. White squares correspond to models that failed to converge.
Each panel represents a variation in input physics, described in Section \ref{sec:method}.
The middle panel shows our fiducial physics assumptions; the bottom and top panel show variations with respectively decreased and increased wind mass-loss rates; and the left and right panels show variations with decreased ($D_{ES} = 1$) and increased ($D_{ES} = 4$) mixing efficiency respectively. In each panel, a black line shows the edge of the chemically homogeneous window for the fiducial input physics assumption.
}
\label{fig: CHE_window_all}
\end{figure*}

For each input physics variation explored, we map out which initial masses and initial periods lead to chemically homogeneous evolution. 
We show the resulting `window of chemically homogeneous evolution' for each model grid in Figure~\ref{fig: CHE_window_all}.
Models that evolve chemically homogeneously are shown in blue. Different shades of blue indicate if and when the model first experienced RLOF through the L1 Lagrangian point. Non-chemically homogeneously evolving models are colored in grey, while failed models are left white. 
The center panel shows our fiducial input physics assumption. 
The black line shows the upper boundary for CHE from the fiducial model grid. For comparison, we show the fiducial boundary over each of the model grid panels.

It is interesting to know if our models experience any mass transfer. 
The shades of blue indicate the moment of first mass transfer; the darkest blue indicates RLOF at ZAMS, medium blue indicates RLOF post-ZAMS, and the lightest blue shows models that do not experience any RLOF before core-helium ignition (when we end our simulations).
The radii of pre-main sequence stars are generally larger than ZAMS radii in single-star models, which could lead to the merger of the two objects before even reaching the main sequence. However, it is not known whether stars that form in a close binary have the same properties as single stars, so whether they would merge is an open question.

The left-most and right-most panel show the models with decreased and increased mixing efficiencies, 
respectively. 
We see that increasing the mixing efficiency expands the window for CHE evolution: i.e. at constant initial mass, larger initial orbital periods result in CHE. Similarly, decreasing mixing closes the window, and only systems with shorter initial periods exhibit CHE.
We also find that more/less systems experience mass transfer in the decreased/increased mixing grid. We attribute this to more efficient mixing keeping the stars more compact. 
Additionally, the increased mixing grid produces CHE at larger orbital separations, causing them to be less likely to undergo mass transfer.

Our reduced and fiducial mixing models correspond to the standard and enhanced grid from \cite{Hastings+2020}, respectively. However, our models extend to much longer periods (5 days instead of 2 days) and higher masses ($\log (M/M_\odot)=2.45$, or $M=281\,M_\odot$, instead of $\log (M/M_\odot)=2$, or $M=100\,M_\odot$). We can therefore only compare the bottom left of our left-most and center panels to the first column of their Fig. 7. 
We find that the window for CHE is slightly larger in our study. E.g., for $\log m_0 = 1.9$, all models up to initial periods of 2 days evolve homogeneously, while \cite{Hastings+2020} find that in their standard grid (our reduced mixing grid) CHE ends at periods wider than $\sim1.75$ days. We believe this to be a result of different \texttt{MESA} versions (\citealt{Hastings+2020} uses \texttt{MESA} r10398, which was prior to the r11532 update's changed treatment of rotating stars: see \citealt{Paxton+19}).

In the upper panel of Figure \ref{fig: CHE_window_all}, we see the effect of increasing the stellar winds by a factor of three. Comparing this panel to the fiducial, we see that increasing the mass loss rate widens the window for CHE drastically for higher mass systems. For these systems, the mass loss rates are so high that the remaining hydrogen-rich envelope is stripped off, leaving a homogeneous core \citep[e.g.][]{Brott+2011, Kohler+2015}. For initial masses above $m_0 > 160\Msun$ we find all models appear chemically homogeneous, irrespective of their initial orbital period. In contrast, for lower mass systems we find that increasing the wind mass-loss rates reduces the window for CHE. This can be understood as a result of the increased mass loss causing the binary to widen and the stars to spin down. This spin down reduces Eddington-Sweet circulations and hence prevents CHE. 

In the bottom panel, we see the effect of decreasing the wind mass-loss rates by a factor of three. The chemically homogeneous window does not significantly change from the fiducial, although we observe a slight widening around $m_0 \approx 150\Msun$. 
Since models with lower wind mass-loss rates widen less, the tidal effects imposing synchronous rotation cause them to remain fast-spinning. 
Additionally, more/fewer models experience mass transfer in the decreased/increased wind mass-loss grid. This is a combined effect of the wind widening and envelope stripping. The decreased wind mass-loss models widen much less, and, so, if any given star increases in radius, it is more likely to overflow in a tight binary as opposed to a wide one. Additionally, if stars are stripped less, as in the decreased mass loss grid, then they have more mass and correspondingly larger radii.

\section{Finding the Best Fitting Model for HD~5980}
\label{sec: best fit for HD5980}

\subsection{Bayesian Fit Procedure}
\label{sec:bayesian_fit}

Bayes Theorem states: the probability of a model given the data, $\rm P(M|D)$, is equal to the probability of the data given the model, $\rm P(D|M)$, times the probability of that model, also known as the prior, $\rm P(M)$ divided by the probability of the data, P(D),
\begin{equation}
    \rm P(M|D) = \frac{\rm P(D|M) \cdot P(M)}{\rm P(D)}.
\end{equation}
We describe each of these terms below. 

\paragraph{Probability of the data given the model} 
We estimate $\rm P(D|M)$ using the sum of the square differences between observed data and the corresponding model predictions, scaled by the allowed error for each parameter squared.
We define 
\begin{equation}
    \rm P(D|M) \equiv \exp \left( - \sum_i \frac{\left(\rm D_i - \rm M_i\right)^2}{\rm \sigma_i^2} \right),
\end{equation}
where $\rm D_i$, $\rm M_i$, and $\rm \sigma_i$ respectively represent each observed and corresponding predicted parameter $i$ and its allowed error. 

We consider seven observables, namely, the orbital period $\rm P_\mathrm{orb}$, the masses of both components, $\rm M_{\rm A}$ and $\rm M_{\rm B}$, their luminosities, $\rm L_{\rm A}$ and $\rm L_{\rm B}$ and the surface hydrogen mass fractions, $\rm X_{\rm H_A}$ and $\rm X_{\rm H_B}$. We do not use temperature, as we believe the observed value of temperature is a lower bound (see Section \ref{sec:limitations}).
We adopt the standard deviation of each parameter for its allowed error, $\sigma$, except for the case of orbital period. 
The observed orbital period is determined very precisely \citep[][see Table \ref{tab:params}]{Sterken+1997, Koenigsberger+2014}. The error in period will therefore be dominated by the rough spacing of our model grid. With this in mind, we take the uncertainty on the orbital period to be 2 days, corresponding to the period spacing of models at termination. This is approximately $10\%$ of the measured period, which is consistent with the order of magnitude of the uncertainties on the other observables.

\paragraph{Prior probability} For $\rm P(M)$, we combine the individual prior probabilities of our free parameters: the initial mass, $m_0$, initial orbital period, $p_{\rm orb,0}$, and age, $t$. 
\begin{equation}
    \mathrm{P(M)} \equiv \mathrm{P}(m_0) \cdot \mathrm{P} (p_{\mathrm{orb,0}}) \cdot \mathrm{P} (t).\footnote{Note that we treat all priors as independent, which may not be the case.}
\end{equation}
We assume a Salpeter Initial Mass Function \citep[IMF,][]{Salpeter1955}:
\begin{equation}
    \frac{dn}{d\log m} \propto m^{-1.35}.
\end{equation}
To get $\mathrm{P}(m_0)$, we integrate between the logarithmic halfway points between each consecutive initial mass in our model grid. 
We assume systems form following Opik's Law in period
\citep{Opik1924}:
\begin{equation}
    \frac{dn}{dp_{\rm orb, 0}} \propto  p_{\rm orb, 0} ^{-1}.
\end{equation}
To get $\mathrm{P}(p_{\rm orb, 0})$, we integrate between the linear halfway points between each consecutive initial period.
We further assume a constant rate of star formation such that
\begin{equation}
     \frac{dn}{dt} \propto \rm t^0.
\end{equation}
Thus, P$(t)$ behaves such that each step in the simulation is given a weight equal to the length of its timestep.

\paragraph{Probability of the data} Lastly, $\rm P(D)$ is a constant for all models. We treat it as a normalization constant obtained by calculating the value from marginalizing across all model timesteps in all five of the model grids.

For each evolutionary track (specified by the initial period and initial component mass) and for each time step on that track (specified by the age), \texttt{MESA} gives us a prediction for the observed parameters we wish to compare to. 
We determine P(M$_{\rm snapshot}|$D), which refers to the probability given the data for a specific timestep. 
We then marginalize over each model run by summing over all P(M$_{\rm snapshot}|$D) in an evolutionary track to get P(M$_{\rm track}|$D), the total probability of the evolutionary track given the data.

Finally, we marginalize over each model grid by summing over all evolutionary tracks within the grid to calculate its total (relative) likelihood, $\cal{L}$. We use this quantity to compute the Bayes factor (see Table \ref{tab:all_mod_params}) by dividing this quantity for each variation of the input physics by the quantity from the fiducial grid.

\subsection{Result Obtained from the Fiducial Model Grid} \label{sec:results-fid}

\begin{figure}
\centering
\includegraphics[width=0.49\textwidth]{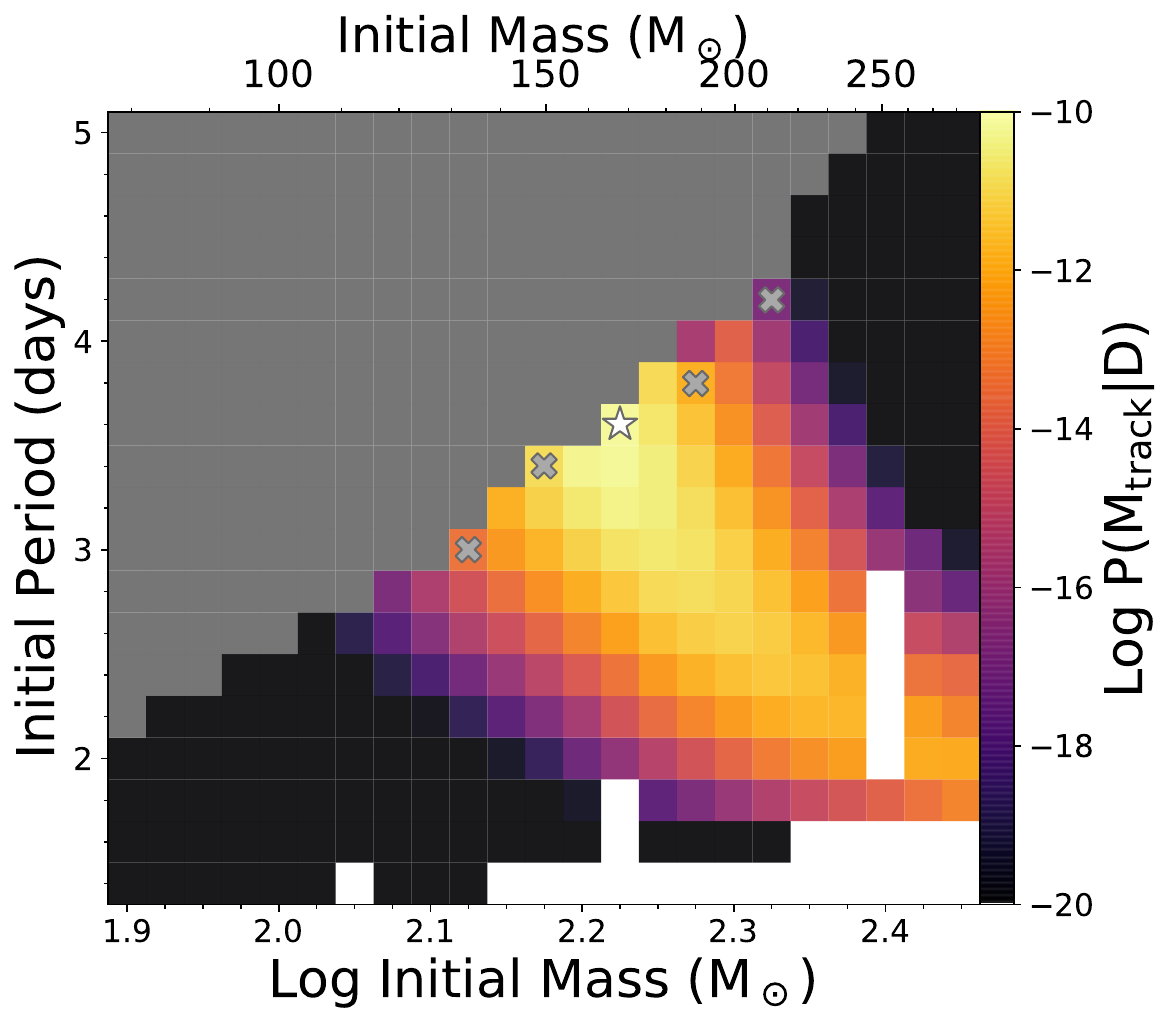}
\caption{The probability of each model run/track in our fiducial model grid given the data as a function of the initial period and component mass, P(M$_{\rm track}|$D). The most probable model is marked with a white star. Gray crosses mark the example models used in Figure \ref{fig: evo_tracks} alongside the most probable model. Grey squares are models which do not evolve chemically homogeneously. White squares show models which failed to converge.}
\label{fig: 2D_prob}
\end{figure}

Using the fit procedure explained in Section~\ref{sec:bayesian_fit}, we compare our models against the observed data of HD\,5980. We compute P(M$_{\rm track}|$D) for each evolutionary track of a binary system with specific initial parameters. The results for our fiducial model grid are shown in Figure~\ref{fig: 2D_prob}. The white star indicates the evolutionary track that has the highest probability given the data, or, simply put, our best fitting model in this grid. 

We draw attention to the diagonal trend in Figure~\ref{fig: 2D_prob}. The best fitting models follow a trend whereby more massive systems lead to better solutions at shorter orbital periods. This is due to the requirement to match the present-day orbital period of about 19 days and mass of about 60\Msun. When starting with a shorter orbital period, more mass loss is needed to widen the orbit sufficiently to match that of the present-day; the initial mass must thus be correspondingly higher to result in the present-day mass.

\begin{figure}
\centering
\includegraphics[width=0.47\textwidth]{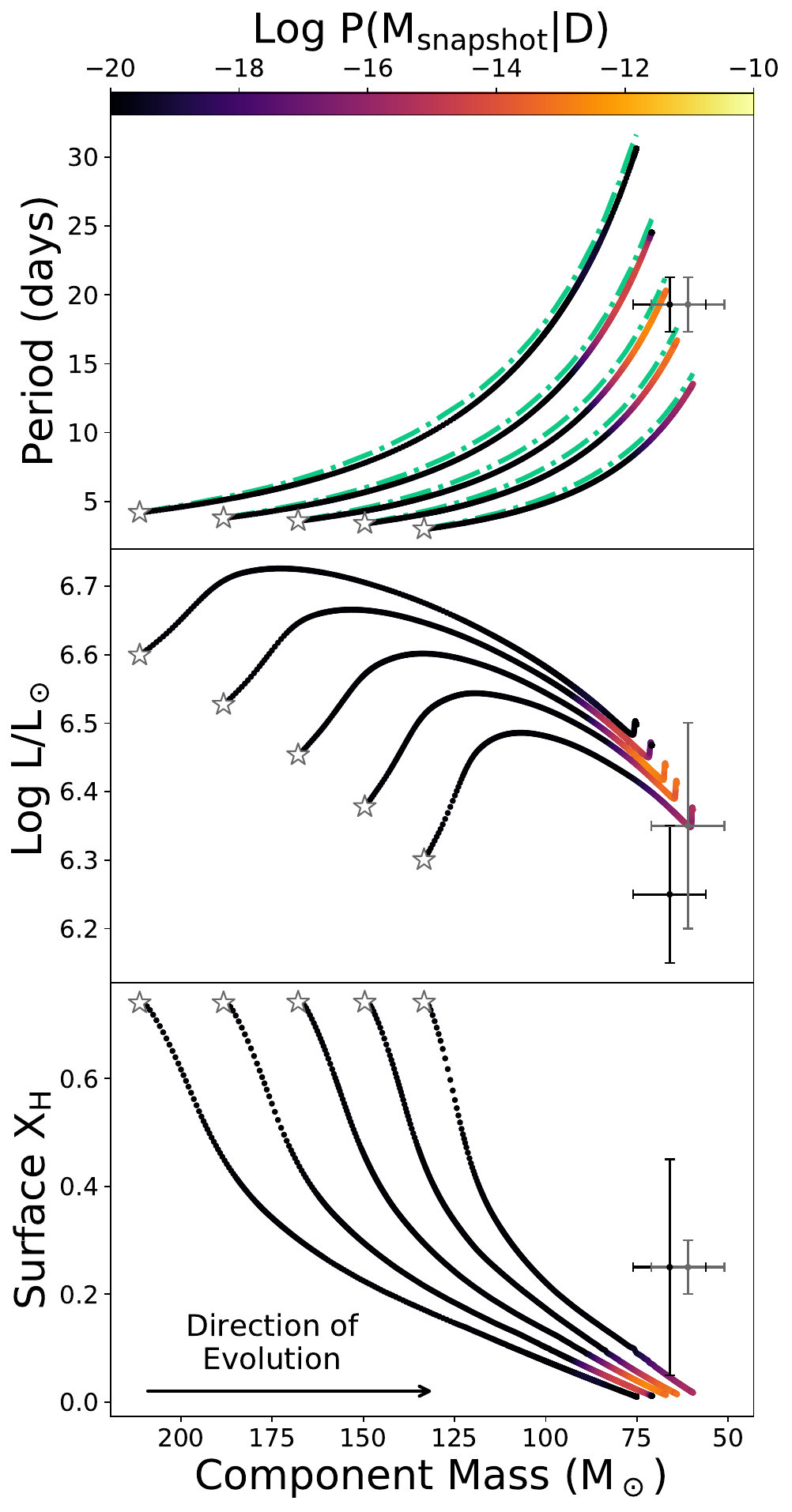}
\caption{Evolutionary tracks of the orbital period, luminosity, and surface hydrogen mass fraction as a function of the decreasing mass. Models start at the star symbol and evolve in the direction indicated. The color indicates P(M$_\mathrm{snapshot}|$D), which is the probability of a given snapshot in time along an evolutionary track given the data. Green dashed-dotted lines in the top panel show the analytical solution as discussed in the text. Observed data for HD\,5980 A (gray) and B (black) are plotted with their corresponding error bars.}
\label{fig: evo_tracks}
\end{figure} 

To illustrate the behavior of the evolutionary tracks, we select five example tracks marked with gray crosses and a white star in Figure~\ref{fig: 2D_prob} and plot their periods, luminosities, and hydrogen abundances as a function of component mass in Figure~\ref{fig: evo_tracks}. Here, we use the component mass as a tracer of the evolution, and the passage of time is tied to evolution from higher to lower masses.
See also Appendix \ref{app:comparison_tracks} for a comparison of evolutionary tracks between all variations of the input physics. 

The upper panel shows how the orbit widens as the mass decreases. This is expected for mass loss in the form of fast stellar winds. 
To show that the wind mass-loss can explain the bulk of the orbital widening, we show the analytical solution for orbital widening in dashed lines \citep[the so-called Jeans mode approximation, see, e.g.,][]{Soberman+1997}. Using conservation of angular momentum and Kepler's law one can derive that the factor by which the orbital period $P_{\rm orb}$ widens depends on the factor by which the total mass changes as,
\begin{equation}
\frac{P_{\rm orb}}{P_{\rm orb, 0}} = \left(\frac{M_{\rm tot}}{M_{\rm tot, 0}}\right)^{-2}. \label{eq:jeans}
\end{equation}
We see that our \texttt{MESA} evolutionary tracks closely follow the analytical solution for Jeans mode mass loss. The remaining discrepancy can be explained by stellar spins and tides, which are not included in Eq. \ref{eq:jeans}. 

We find that the best fitting solutions are near the end of the evolutionary tracks. 
Since our initial models typically start with period of 2-4 days, we need an increase in period by a factor between 5-10 to match the present-day orbital period of about 20 days. Eq.~\ref{eq:jeans} explains why the initial masses of our best fitting models are thus in the range of $2-3 \times$ higher than the present day observed mass, e.g.\ $120-180 \Msun$. \\

In the middle panel of Figure \ref{fig: evo_tracks} we show how luminosity evolves as a function of star mass. For each track, we initially observe an increase in the luminosity as the system evolves. This is because of the increased helium content as the star burns hydrogen. Later in the evolution, we see a turnover where the luminosity begins to decrease. This is the result of high mass loss. The best fitting models tend to be somewhat over luminous, although they are still well within 2-$\sigma$ of the observations.

Finally, in the bottom panel, we show the evolution of the surface hydrogen abundance. We note that the best fitting snapshots tend to predict lower surface hydrogen abundances ($X \lesssim 0.1$) than observed (which are $\approx0.25$). We further discuss this discrepancy in Section \ref{sec:limitations}.

\subsubsection{Posteriors for the Model Parameters}
\begin{figure}%[ht]
\centering
\includegraphics[width=0.49\textwidth]{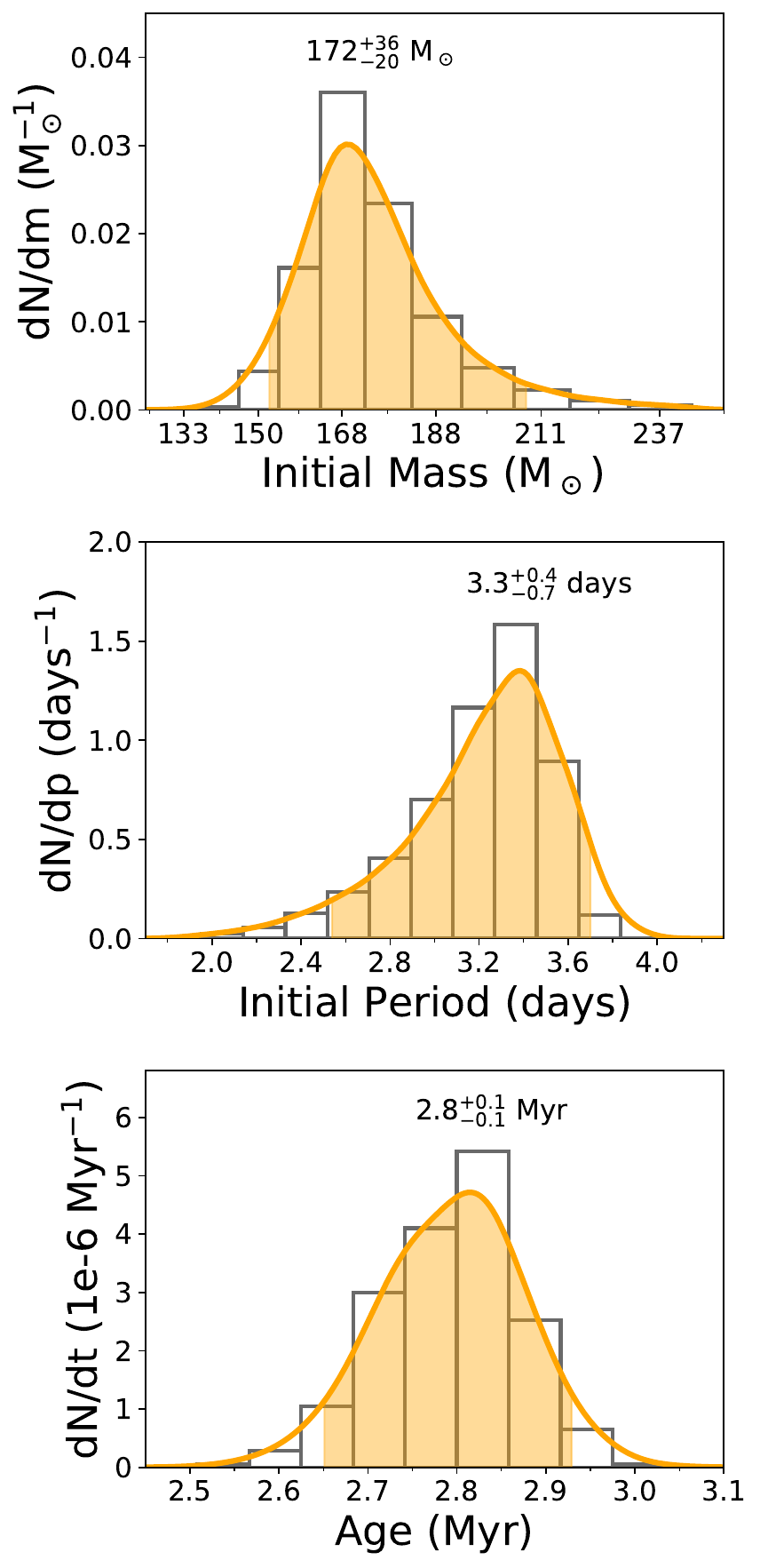}
\caption{Posterior probability distributions of our model parameters: initial component mass, initial orbital period and the age. The median value is displayed above the curve, and the region bounded by the 5th and 95th percentiles are indicated by shading under the curve. In the upper and middle panel, the width of the bins corresponds to spacing in our model grid. In the bottom panel, bins are chosen by eye to be wide enough such that discrete features from our grid do not show up.}
\label{fig:post_model_parameters}
\end{figure}

In Figure~\ref{fig:post_model_parameters} we show the posterior probability density functions for the initial parameters of the binary systems and the age. The orange-shaded regions correspond to the intervals bounded by the 5th and 95th percentiles. 

The median (and 5th/95th percentiles) of the posterior distribution of the initial mass is $172^{+36}_{-20}$ M$_\odot$, as can be seen in the top panel. This is high compared to most known massive stars, but stars with such masses have been claimed to exist in the Large Magellanic Cloud. In particular, such massive stars exist in the center of the massive star cluster R136 \citep[e.g.][]{Crowther+2010, Crowther+2016, Brands+2022} and also nearby \citep{Bestenlehner+2011, Renzo+2019b}, although we note that HD\,5980 has been observed on the outskirts of its host cluster, NGC\,346.
We stress that the stellar evolutionary models for such high mass stars are still very uncertain and this result should be taken with appropriate caution.

The median of the initial orbital period posterior distribution is $3.3^{+0.4}_{-0.7}$ days, (middle panel). Such short periods are typical for early type stars \citep[e.g.][]{Sana+2012}. This period is also reminiscent of the massive galactic binary WR20a, which contains two hydrogen WR stars in an orbit with a period 3.7 days \citep[see Appendix \ref{app: further CHE} and][]{Rauw+2005,Bonanos+2010}. 

For the age of the system we infer a median value of $2.8^{+0.1}_{-0.1}$ \Myr (bottom panel). This is consistent with the age derived for nearby components of NGC 346, which is approximately $3\pm1$\,Myr (see Table \ref{tab:params}).

\subsection {Variations in Wind Mass-Loss and Chemical Mixing \label{physics_variations}}

\begin{figure*}
\centering
\includegraphics[width=\linewidth]{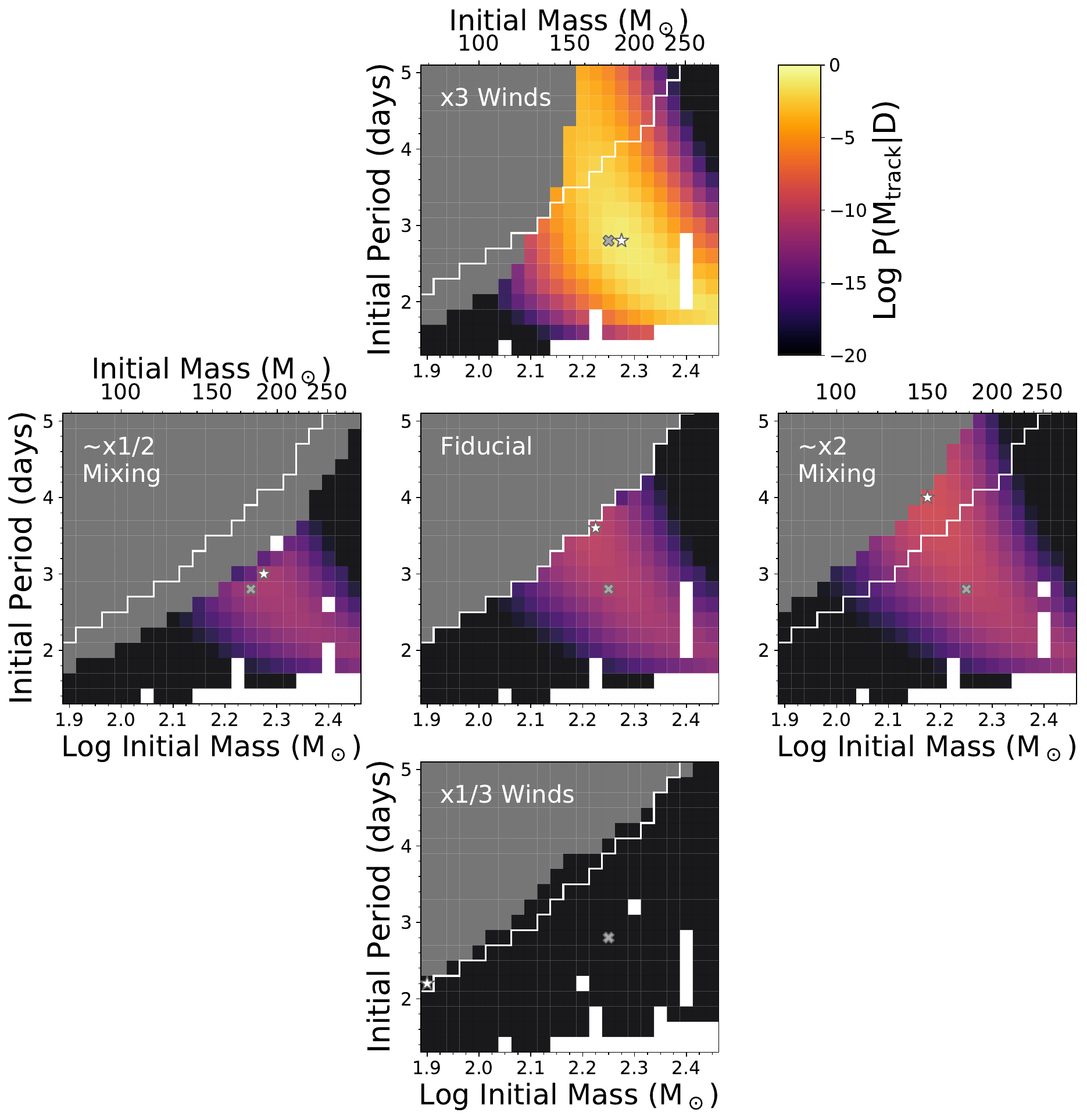}
\caption{Probability maps showing P(M$_{\rm track}|$D) for each of our input physics variations. Panels showing the grids of models for each of the variations of the input physics are laid out as in Figure \ref{fig: CHE_window_all}. In white, we overplot the fiducial boundary for chemically homogeneous evolution for comparison among the input physics variations. The best fitting model in each grid is marked with a white star. Gray crosses show the models which are used for comparison between the input physics variations in Appendix \ref{app:comparison_tracks}.}
\label{fig:phys_variations_2Dpmd}
\end{figure*}

In Figure~\ref{fig:phys_variations_2Dpmd} we show the probability maps for each input physics variation. 
We see the same diagonal band in each panel as for our fiducial grid, indicating a degeneracy between increasing initial mass and decreasing orbital period. The decreased winds panel appears to be the exception, but in this case the probabilities are too low to see the variation due to the chosen colorbar (and the trend still exists). 

Most striking in this figure are the drastically increased probabilities found within the model grid for increased stellar winds, shown in the top panel. 
These high mass-loss evolutionary tracks are the best at recreating the observed parameters of HD\,5980, especially in luminosity and surface hydrogen abundance (see Figure \ref{fig:phys_variations_evo}). For each of the explored input physics variations, we show the evolutionary behaviour in Appendix \ref{app:comparison_tracks}. 

In the bottom panel, where we show models with decreased mass loss, the probabilities are much lower than in the other panels and they fall below our chosen lower bound of the colorbar. We see that with decreased mass loss, it becomes very hard to explain the properties of HD\,5980. As explained above, this is because significant widening is required to match the present-day orbital period of 19.3 days, while the reduced mass loss rates prevent such widening from occurring. These models cannot simultaneously match the masses and periods observed for HD\,5980. Thus, the most likely evolutionary track in this grid is the one with the lowest initial mass and highest initial orbital period.

Comparing the variations in mixing, shown on the left and right panel of Fig.~\ref{fig:phys_variations_2Dpmd}, we see that increased mixing generally increases $\rm P(M_{\rm track}|D)$. This effect is much less significant than the effect of the winds.

For most of the grids, the best fitting evolutionary track (marked with a white star symbol) can be found near the edge of the chemically homogeneous window. These models have the largest initial periods with the lowest allowed initial mass while still experiencing tidal CHE. Wider periods and lower masses are favored by our priors on the initial distributions of orbital periods and stellar masses.  

For an overview of all the posterior distributions of the parameters from our analysis, see Figure~\ref{fig:hist_obs} in Appendix \ref{app:posteriors var}.

\subsubsection{Bayes Factors Between Model Variations} 
\label{sec: bayes_factor}

To compare our different model variations, we compute their Bayes factor relative to the fiducial model grid.
In Table~\ref{tab:all_mod_params}, we provide an overview of the Bayes factors together with the median and 5th/95th percentiles for the initial model parameters. 
We see that the model grid with increased mixing is somewhat preferred over our fiducial model grid. 
% Lieke working on this
However, the increased winds model grid is strongly favored over the fiducial.
The preference of this model can be largely attributed the orbital widening that follows from wind mass loss. Since our models have short initial orbital periods necessary for tidal locking (2-4 days), a large amount of mass loss is needed to reach the present-day observed period of $\sim19\,$days (see also the top panel of Figure~\ref{fig: evo_tracks} and the corresponding discussion in Section \ref{sec:results-fid}). 
For this same reason, the model grid with reduced wind mass-loss is very strongly disfavored: these models fail to sufficiently widen the binary orbit to the observed value. 
As mentioned in Section \ref{sec:method}, winds remain one of the main uncertainties in massive-star models. Our models imply that in order to explain HD\,5980 with tidal CHE, the winds of these very massive stars need to be significantly larger than the optically thin winds of OB stars as described by \cite{Vink+2001}.

\begin{table}[ht]
\centering
\scriptsize
\begin{tabular}{ccccc}
\hline
Input Physics & Bayes Factor & $m_0$ & $p_{\rm orb, 0}$ & $t$ \\ 
$\left[\rm Units\right]$ & $\log_{10} \cal{L}/\cal{L}_{\rm fiducial}$ & [$\Msun$] & [days] & [\Myr] \\ 
\hline \hline
Fiducial & $0$ & $172^{+36}_{-20}$ & $3.3^{+0.4}_{-0.7}$ & $2.8^{+0.1}_{-0.1}$ \\ 
\hline
Reduced Mixing & $-1.15$ & $201^{+51}_{-25}$ & $2.7^{+0.4}_{-0.6}$ & $2.7^{+0.1}_{-0.1}$ \\ 
Increased Mixing & $1.20$ & $159^{+25}_{-17}$ & $3.7^{+0.6}_{-0.8}$ & $2.9^{+0.1}_{-0.1}$ \\  \hline
Reduced Winds & $-20$ & $80^{+9}_{-3}$ & $2.2^{+0.2}_{-0.2}$ & $3.4^{+0.2}_{-0.3}$ \\ 
Increased Winds & $9.29$ & $199^{+66}_{-34}$ & $2.6^{+0.8}_{-0.7}$ & $2.5^{+0.1}_{-0.1}$ \\ 

\end{tabular}
\caption{The total likelihood $\cal{L}$ for each model grid given the data is given relative the total likelihood for our fiducial model $\cal{L_{\rm fid}}$. Positive values indicate input physics variations that are preferred over the fiducial assumptions. This shows the increased mixing and increased winds variations are favored over our fiducial input physics assumption. We also list the corresponding median and 5th/95th percentiles for the initial component mass, $m_0$, initial orbital period, $p_{\rm orb,0}$, and the age of the best fitting model, $t$, for HD\,5980.}
\label{tab:all_mod_params}
\end{table}

In summary, we find that enhanced mixing and, especially, enhanced wind mass-loss help to explain the observed properties of HD\,5980. 
It is worth noting that within the enhanced wind mass-loss grid, the stellar behavior tends to resemble that of wind-stripped stars rather than stars undergoing homogeneous evolution due to rotational mixing. 
While the initial homogeneous evolution is driven by tidal interactions, the stellar winds pick up as we switch to the \cite{Hamann+1995} prescription when the surface He abundance increases to $Y_{s}>0.7$. 
Consequently, further homogeneous evolution occurs, at least in part, because stellar winds strip away the star's envelope; these outcomes rely heavily on the adopted wind mass-loss prescription.

\section{Further Evolution and Final Fate} 
\label{sec:future evol}

So far, we have focused on uncovering the possible progenitor system of the present-day HD\,5980 A\&B binary. Equally interesting is its possible future evolution. 
In this section, we investigate the future fate of HD\,5980, both under the enhanced wind mass-loss and fiducial assumptions.

\subsection{Binary Black Hole Formation and Potential Pair Instability Pulsations}
\label{sec:final_fate}

We use \texttt{MESA} to further evolve our best fitting models from the enhanced wind mass-loss grid ($m_0 \approx 188\Msun$, $p_0 = 2.8 \,\rm{days}$) and the fiducial grid ($m_0 \approx 170\Msun$, $p_0 = 3.6 \,\rm{days}$) until core collapse. For these calculations we apply the method of \cite{Marchant+2019}, \cite{Farmer+2019a}, and \cite{Renzo+2020b}.  

After completing their advanced burning stages, the stars lose most of their mass via stellar winds, leaving behind massive CO cores with a thin helium layer on top.

For the preferred, best-fitting enhanced wind mass-loss model, we find the final CO core mass at C-ignition to be about $19\Msun$, for both components. A CO core with this mass is expected to experience core collapse directly into a black hole, without losing further mass due to pair production. Thus, the final fate of this model is a binary black hole with masses of about $19\Msun$.

For our fiducial model, we find the CO core masses at C-ignition to be about $37\Msun$, which places them just above the lower bound to experience a pair-pulsational supernova \citep[PPISN,][]{Fowler+1964,Barkat+1967,Rakavy+1967}. 
The CO-core loses approximately $0.1\Msun$ in pulses before finally collapsing into a black hole.  

This would result in a black hole binary with two black holes of about $36.5\Msun$ \cite[ignoring any further mass loss during the collapse to a black hole, e.g.,][]{Marchant+2019}.
This implies a chirp mass of about $32\Msun$, which is slightly above the location of the observed high-mass feature in the mass distribution of merging binary black holes as detected by LIGO and Virgo \citep[which is currently constrained to lie at an approximate chirp mass of $27.9^{1.9}_{-1.8M}\Msun$,][]{GWTC-3_popPaper}.  
This feature has often been attributed to the pulsational pair-instability pile up \citep[e.g.,][]{TalbotThrane2018,Farr+2019,Stevenson+19}, which would agree with our result presented here. 
However, recent updates of the tabulated $^{12}\rm{C}(\alpha,\gamma)^{16}O$ reaction rates have pushed the lower boundary of pair-pulsational supernova, and the corresponding expected location of a pile-up in the remnant mass distribution, to higher masses \citep{Farmer+2020,Mehta+2022,Farag+2022,Shen+2023, Hendriks+2023}.
More updated prescriptions would therefore not result in a PPISN, but in direct core-collapse supernova. Given the low mass loss during the pulse, our final CO-core mass would still lead to a comparable black hole mass via core collapse, and nonetheless lie interestingly close to this observed feature.

In summary, we expect that HD\,5980 A\&B will result in CO-core masses $\sim19-37\Msun$, producing a binary black hole with approximately those same masses. Our preferred model from the enhanced wind mass-loss grid falls within the regime for core collapse. When we instead follow the fiducial model, we find CO-core masses that are either on the edge, or just too low-mass, to experience PPISN; and which produce binary black holes with masses that coincide with the high-mass feature in the mass distribution of merging binary black holes (chirp mass $\sim 32\Msun$).

The enhanced wind mass-loss and fiducial models widen from their ``present-day" state due to wind mass loss in their later evolutionary stages. We find final orbital periods of about $220$\,days and $55$\,days, respectively. These systems are both too wide to merge within the age of the universe due to gravitational wave driven in-spiraling \citep[][]{Peters1964}; however, we discuss a possible merger scenario in the following section.

\subsection{Speculation on the Role of the Third and Fourth Companions}
\label{sec:secular_evolution}
\newcommand\ahc[1]{{\color{red}\bf#1}}
HD\,5980 A\&B is thought to be part of a hierarchical 2+2 quadruple system. This may have interesting implications for the final fate of the binary black hole \citep[see, e.g.,][]{Antonini+2017,Fragione+2019, Vynatheya+2022,VignaGomez+2022}. In this section, we speculate on the ultimate fate of HD\,5980 A\&B.

Little is know about the nature of the triple and quadruple components and the parameters that characterize their orbits. \citet{Koenigsberger+2014} quote an orbital period of 96.56 days and an eccentricity $(e=0.82)$ for third star, star C, and its unseen fourth companion, which we refer to as component D. Star C is estimated to have a mass of about $30 \Msun$ \citep{Shenar+2016}. The mass and nature of component D is not known. Neither is the mutual orbit of the A\&B and C\&D binaries known, and it may in fact be simply a chance alignment. However, given that massive stars often occur in higher order multiple systems \citep[e.g.][]{Sana+2014,2017ApJS..230...15M,Offner+2023}, we will speculate on possible consequences of a quadruple nature in the remainder of this section.  

If system A\&B and system C\&D indeed form a bound 2+2 quadruple system, we can estimate the minimum mutual orbital period (i.e., the outer orbital period) required for dynamical stability. 
Throughout the rest of this section, we will assume a plausible range of masses of 3-40$\Msun$ for the unseen companion D (spanning the parameter space from a lower-mass star to a high-mass compact object). 
With this, we use Eq.\ 90 from \cite{Mardling+2001} on the present day parameters to estimate that the mutual orbital period  has to be a least $\sim500$ days, 
as shorter periods would cause the system to enter the regime for chaotic interactions, likely leading to a (partial) break up of the quadruple. 

Both inner binaries in a 2+2 quadruple system may experience secular evolution due to long-term gravitational torques, leading to eccentricity oscillations (see \citealt{2016ARA&A..54..441N} for a general review on secular evolution in triple systems, though quadruple systems are generally more complicated). The inner binary system with the longer period (containing Star C and its companion D, with a period of 96.56 days) would experience secular eccentricity oscillations on a shorter time scale than the inner binary with the shorter period (containing stars A and B, with a period of 19.3 days), since the secular time-scale generally scales as $P_\mathrm{out}^2/P_\mathrm{in}$. This difference may explain why the C\&D system has a higher eccentricity ($e=0.82$), and thus possibly hint that the system is indeed presently undergoing secular evolution. We note also that secular evolution, in addition to eruptive mass loss episodes, may have contributed to the eccentricity of the A\&B system. 

Generally, hierarchical quadruples are susceptible to chaotic secular evolution, meaning the eccentricity evolution can be highly complex and much higher eccentricities can be reached for more orthogonal inclinations \citep[e.g.,][]{2013MNRAS.435..943P,Hamers+2017,2018MNRAS.474.3547G}. Whether or not this type of chaotic secular evolution occurs can be characterized by the ratio of the secular timescales of both orbit pairs in the 2+2 system. In \citet[][Eq.\ 32]{Hamers+2017}, this ratio is defined as $\beta$, where $\beta$ values close to $1$ experience significant eccentricity boosts. 
Based on the current state of the HD\,5980 system, reasonable values of $\beta$ (here defined as the ratio of the secular time-scale of the A\&B orbit to the outer orbit, to the secular time-scale of the C\&D orbit to the outer orbit) are in the range $5 \lesssim \beta \lesssim 15$.
This means that the current system is probably not in the regime for chaotic secular interaction \citep[see e.g.\ Fig~2 of][]{Hamers+2017}.

However, if the A\&B components evolve to $37\Msun$ BHs in a 55-day orbit (as we suggest in Section~\ref{sec:final_fate} for our fiducial model), then the $\beta$ ratios get much closer to unity, namely $0.9 \lesssim \beta \lesssim 5.1$.
If the components instead evolve to $19\Msun$ BHs in a 220-day orbit (as suggested for our enhanced winds model), then the $\beta$ ratios are even closer to unity, falling in the range $0.7 \lesssim \beta \lesssim 3.1$. In the latter case, the inner period of $220$\,days may even be wide enough relative to the (unknown) outer orbital period that the system is brought into the chaotic regime.

This could suggest that eccentricity excitation in A\&B’s orbit, if not efficient right now, could possibly become very significant through secular chaotic evolution by the time the BHs form, significantly accelerating their GW-driven merger. This argument is dependent, however, on short-range forces not quenching secular evolution. The relativistic apsidal motion timescale in the BH A\&B orbit is calculated to be $\sim 10^5 - 10^6$ yr. The secular time-scale in orbit A\&B is similar or shorter than the apsidal-motion timescale for outer orbits shorter than approximately 50 yr, which is plausible given that the minimum current outer orbital period for stability is $\sim 500$ days, or 1.4 yr.  
Therefore, relativistic quenching should be unimportant and not impede the coalescence of the BHs, unless the outer orbit is fairly wide.

Above, we assumed that the components in the C\&D orbit do not merge before the A\&B stars successfully evolve to BHs. It is possible, however, that the C\&D components merge at an earlier time due to secularly-driven high eccentricity. In this scenario and depending on the outer orbit, the merger remnant could potentially fill its Roche-lobe around the A\&B BH binary system as it evolves \citep[see, e.g.][]{2014MNRAS.438.1909D,2021MNRAS.500.1921G,Hamers+2022,Merle+2022}. This could in turn lead to either stable mass transfer or common envelope evolution; in the latter case, the BHs could merge during the process in a gas-rich environment, and potentially produce EM counterpart observations. 

Regardless of the uncertain final fate of the HD\,5980 system itself, we conclude that systems like it can be very interesting as progenitors for gravitational wave events, which may even be prompt and associated with an EM counterpart.

\section{Model Limitations and Alternative Evolutionary Explanations}
\label{sec:limitations}

We have explored a possible explanation for the HD\,5980 A\&B system through a pathway of chemically homogeneous evolution starting from a tight binary system (tidal CHE). 
In general, the solution we found involves very high initial masses, for which the stellar evolution is highly uncertain. 
Moreover, although we can satisfactorily explain the present-day period, masses, and luminosity, we were unable to explain several other observed properties of HD\,5980.

Firstly, our fiducial, reduced, and enhanced mixing models are at tension with observed hydrogen abundances (see Figures \ref{fig:phys_variations_evo} and \ref{fig:hist_obs}). Our enhanced mass-loss models do reproduce the surface hydrogen abundance within two standard deviations on the primary and one standard deviation on the secondary.  

Second, HD\,5980 shows evidence for eruptive mass loss episodes \citep[see, e.g.,][]{Bateson+1994, Barba+1995, Koenigsberger+1995, Sterken+1997}. We do not account for this in our models, given that the nature of these eruptions is still unknown. We have also not addressed the eccentricity, though we note it is possible the effect of secular evolution may amplify any small eccentricity initially contained within, or later introduced to the system, 
e.g. through significant mass-eruptions. It is possible that these mass eruptions themselves induce eccentricity, similarly to the Blaauw kick from spherically symmetric supernova ejecta.
However, we expect the latter effect to be small, assuming that these previous eruptions were similar to the 1994 event in which approximately $10^{-3}\Msun$ was ejected in total (which is much smaller than is typically ejected in a supernova event).

We have not fit our models to observed constraints on the surface temperature due to significant uncertainties in its value. HD\,5980 A\&B are experiencing high rates of mass-loss, effectively embedding the stars within an optically-thick wind. We believe the observationally derived value of $T_\mathrm{eff}$ will measure the temperature of the wind and is thus significantly lower than the temperature of the photosphere as computed by \texttt{MESA}. 
Indeed, when modelling HD\,5980, \citet{Georgiev+2011} define the photosphere at the sonic point radius $R_{s}$, and find $T_s(R_s)=60$\,kK. The effective temperature of their model is defined as the temperature at the radius where $\tau=2/3$, and this point lies above the photosphere in the optically-think wind region. They find $T_\mathrm{eff}=43$\,kK, which is consistent with the findings from e.g., \cite{Shenar+2016} of $T_\mathrm{eff}\approx45$\,kK.
Because the sonic point ($R_s=19.6\Rsun$) lies closer to the hydrostatic stellar radius \citep[e.g.,][]{Grafener+Vink2013} we expect $T(R_s)$ from \citet{Georgiev+2011} to be closer to the photospheric temperature from our \texttt{MESA} model. 
We do indeed find, as can be seen in Figure \ref{fig:hist_T}, that our best fitting models overpredict the observed $T_\mathrm{eff}=43$\,kK, but are consistent with the inferred sonic-point temperature from \cite{Georgiev+2011}. This is particularly encouraging given that we did not fit for the temperature.  \\

The results obtained in this study furthermore rely crucially on two assumptions:
1) the mixing is dominated by meridional circulation, which depends strongly on the rotational velocity, and 2) synchronous rotation is imposed throughout the time over which the star is evolved.
Synchronous rotation implies that the star rotates as a rigid body. Thus, there is no differential rotation and no shear instabilities can be triggered. Standard models predict that the core contracts as hydrogen is depleted and, due to the conservation of angular momentum, it speeds up. At the same time, as the outer layers expand, they slow down. Hence, classically, a significant differential rotation structure develops before RLOF occurs. This can lead to significantly more mixing than if synchronous rotation is imposed at all times \citep{Song+2013}. Furthermore, true synchronous rotation can only occur in a circular orbit, and HD\,5980 has significant eccentricity. Thus, its interior layers are continuously subjected to shearing motions which could lead to enhanced mixing.

Also important is the effect of tides, which are highly uncertain. Both arguments for less efficient tides \citep[e.g.,][]{Yoon+2010,Qin+2018,TownsendSun2023,Sciarini2024}, and more-efficient tides \citep{Witte+1999, Ma+Fuller2023} have been made. 
Since the stars in our CHE models are on such close orbits, slightly less efficient tides will not significantly impact their main-sequence evolution (since they would still synchronize before most of the mixing takes place). Hence, we do not expect the window of CHE to be significantly affected. 
Additionally, tides are important in deciding whether the stars remain locked as they evolve into Wolf-Rayet stars, and thus affect the spin of the final remnants \citep[see e.g., ][]{Qin+2018, Bavera+2020}.

\subsection{Alternative Evolutionary Explanations} \label{ss: alternatives}

In this work, we have focused on the chemically homogeneous evolutionary pathway to HD\,5980 due to tidal interactions from ZAMS. 
However, we recognize that in both our fiducial and enhanced mixing grids, the maximum of the posterior distribution lies at the boundary of the chemically homogeneous window. This could imply that the `best' solution lies beyond our definition of CHE. 
Below, we discuss some of the plausible alternative formation pathways.

\paragraph{Post mass-transferring binary}
HD\,5980 A\&B could have started as a wider binary system that experienced a mass transfer phase, causing the orbit to shrink to its current separation. As mentioned in Section \ref{sec:intro}, this is thought to be unlikely because \textit{both} stars look relatively evolved \citep{Koenigsberger+2014}. 
However, the unusual properties of this system (e.g., massive stars, high eccentricity, LBV-like outbursts, etc) may open the door to uncertainties which allow for an unusual evolution in the standard pathway. \cite{Koenigsberger+2021} suggested that the mixing efficiency can be enhanced in binary systems that rotate asynchronously --- this would allow for relatively lower, and more believable, initial masses for the system. 
Additionally, the eccentricity of the system implies the presence of steep velocity gradients in the shearing flows around the time of periastron passage, which may lead to enhanced mass-loss rates around periastron passage \citep{Moreno+2011}. It is not clear how such periastron passage events may impact the overall mass-loss properties, which is further complicated by LBV-type outbursts; hence the evolutionary path is subject to uncertainty.

It is non-trivial to obtain converged models of high mass stars that evolve through mass transfer, but we note that including such peculiar effects into the post-mass transfer scenario might constitute one of the most interesting cases for follow-up studies.

\paragraph{Accretion-induced CHE}
HD\,5980 A\&B may have started as a wider binary that experienced a mass transfer phase, causing the orbit to shrink and spinning up the accreting star, which would then evolve chemically homogeneously. Recent work by \cite{Ghodla+2023}, building on the work done by, e.g., \cite{Wilson+Stothers+75}, \cite{Packet1981}, and \cite{Cantiello+2007}, proposed that the majority of chemically homogeneously evolving stars may be derived from this channel. 
This pathway could explain the relatively evolved appearance of \textit{both} stars in HD\,5980, assuming that we observe the systems post-mass transfer. It would require a stable mass transfer phase from a relatively unequal mass ratio system ($q_{\rm ZMAS}\approx 0.5$), and inefficient mass transfer (less than $10\%$ of the donated mass accreted). Both these conditions are plausible and we highlight this as a possible, but unexplored, pathway in the context of HD\,5980.

\paragraph{Birth-spin CHE}
This is the evolutionary pathway as suggested by \cite{Koenigsberger+2014}, and \cite{Schootemeijer+Langer2018} (see Section \ref{sec:obs HD5980}). 
This pathway can explain the relatively wide observed orbital period of HD\,5980 without the need for heavy mass loss (as is required for the tidal CHE models explored in this work).
However, birth-spin distribution of massive stars is highly uncertain \citep[see e.g.,][and references therein]{de-Mink+2013}.

\section{Conclusion \& Summary} 
\label{sec:conclusion}

We have explored the hypothesis that the massive binary star system HD\,5980 A\&B is a system that formed through tidally induced chemically homogeneous evolution starting from a (near-) contact binary system.
We show that the tidal CHE scenario can explain the short orbital period, present-day masses, luminosities, and sonic-point temperatures. Under this interpretation, we find that the progenitors were at least 150\Msun at zero age and orbiting each other with a period of about 3 days. 
Our increased wind mass loss models are most favored to explain HD\,5980, since a lot of mass loss from winds is needed to bring the binary from a tidally locked configuration at birth ($P<4$d), to the observed orbital period of $\sim20$d. Our models imply that, if the tidal CHE scenario correctly describes HD\,5980, the winds of very massive binary stars are different (much higher) than the commonly assumed optically thin winds
of OB stars from \cite{Vink+2001}. Perhaps this may be explained, in part, by periods of enhanced mass loss corresponding to the observed LBV-like behavior of HD\,5980 A.
We further find only slightly improved fits if we consider enhanced internal mixing. This can be taken as mild support for more efficient mixing within HD\,5980-like CHE stars than is assumed in \citet{Hastings+2020}, which would be in line with the enhanced mixing proposed by \cite{Koenigsberger+2021}.

We discuss the implications of the CHE interpretation.  We find that HD\,5980-like models have nearly completed their hydrogen-burning phase. The present-day masses of about 60\Msun are thus order of magnitude close to the mass of their helium cores. Under fiducial assumptions, these models are candidates for pair(-pulsational) instability supernovae; they result in a binary black hole system with component masses of about $37\Msun$, which closely matches the observed feature in binary black hole masses at 35\Msun \citep[as reported by the LVK collaboration,][]{GWTC-3_popPaper}. Under our preferred enhanced wind mass-loss assumptions, these models instead experience core collapse, producing a binary black hole system with masses of about $19\Msun$. Although the binary black holes produced via forward modelling would be too wide to inspiral due to gravitational wave emission alone, we note that HD\,5980 has been claimed to be a member of a quadruple system ---dynamical interaction with the system's third and fourth components may drive the binary black hole to a merger within a Hubble time. Depending on the behavior of the third and fourth components, this merger may even have an EM counterpart.

We stress that the evolution of such very massive stars is still poorly understood, and hence our models are sensitive to uncertain physical assumptions. Although the CHE interpretation is promising, we note that our current models do not reproduce the observed hydrogen and helium abundances well. We do not exclude alternative formation pathways for HD\,5980. 
Nonetheless, we find this system is a likely binary black hole progenitor, which provides important clues for understanding the evolution of massive binary stars and the formation of possible gravitational-wave sources.

\software{This research made use of \texttt{MESA}: Modules for Experiments in Stellar Astrophysics, r12778 
\citep[\url{http://mesa.sourceforge.net},][]{Paxton+2010,Paxton+2011,Paxton+2013,Paxton+2015,Paxton+2018,Paxton+19}, 
Astropy \citep[\url{http://www.astropy.org},][]{Astropy-Collaboration+2013, Price-Whelan+2018, Astropy2022},
jupyter \citep[\url{https://jupyter.org},][]{jupyter},
matplotlib \citep[\url{https://matplotlib.org},][]{matplotlib},
multimesa (\url{https://github.com/rjfarmer/multimesa}),
numpy \citep[\url{https://numpy.org},][]{numpy}, and
scipy \citep[\url{https://scipy.org},][]{scipy}. 
The \texttt{MESA} inlists used in the creation of the five model grids have been uploaded to zenodo, with DOI 10.5281/zenodo.10594480 (\href{https://zenodo.org/records/10594480}{https://zenodo.org/records/10594480}).}

\begin{acknowledgments}
%People
We are very grateful for Adrian Hamers for discussions on the consequences of the possible quadruple nature of the system. We also thank Ruggero Valli for a helpful discussion on tides, and Tomer Shenar, Pavan Vynatheya, and Jay Gallagher for their valuable comments and insights. Lastly, we would like to acknowledge the anonymous reviewer, whose comments helped to substantially improve the text.
%Grants
This project was funded in part by the Netherlands Organization for Scientific Research (NWO) as part of the Vidi research program BinWaves with project number 639.042.728 and the National Science Foundation under Grant No. (NSF grant number 2009131). GK acknowledges funding support from UNAM DGAPA/PAPIIT  grant IN105723. We further acknowledge the Black Hole Initiative funded by the John Templeton Foundation and the Gordon and Betty Moore Foundation. 
\end{acknowledgments}

\clearpage

\appendix

\section{Further binary candidates for CHE \label{app: further CHE}}
HD 5980 is not the only binary system that has been discussed in the context of CHE. Below, we briefly summarise a (non-exhaustive) list of further candidates.

\paragraph{VFTS\,352} is a massive double-lined O-type spectroscopic binary system in the LMC studied in detail by \cite{Almeida+2015}, who suggested it as a candidate for CHE. VFTS\,352 is one of the strongest candidates for CHE to date, together with HD\,5980 (see Section \ref{sec:obs HD5980}), because of its short orbital period \citep[1.124 days,][]{Almeida+2017}, massive components \citep[$28.63\pm0.30$\Msun and $28.85\pm0.30$\Msun,][]{Almeida+2015}, and rapid rotation \citep[about $290 \mathrm{km\,s^{-1}}$,][]{Ramirez-Agudelo+2015}. Nonetheless, \cite{Abdul-Masih+2019} more recently performed a detailed atmospheric analysis of VFTS\,352 and concluded that even CHE models cannot provide a good fit for this system. Although they find a tidal CHE model that provides an explanation for its location in the Hertzsprung-Russel diagram, they simultaneously over-predict the N and He enrichment with respect to observations.

\paragraph{SMC AB\,8} is an SMC binary containing a WR star ($19^{+3}_{-8}$\Msun) and a companion O-type star ($61^{+14}_{-25}$\Msun) on a 16.6 day period. 
\cite{Shenar+2016} performed spectral analysis of all multiple WR systems in the SMC known at the time. They compare their observed properties to BPASS \citep{Eldridge+2008} models assuming either no mixing or CHE, and found that AB 8 was consistent with the (birth-spin) CHE models. They conclude that CHE can explain its evolutionary state, although it is not a necessary assumption.

\paragraph{R\,145} is an eccentric ($e=0.79$), wide ($P_\mathrm{orb} \approx 160$ days) WR binary in the Large Magellanic Cloud (LMC) with masses $53^{+40}_{-20}$\Msun and $54^{+40}_{-20}$\Msun studied in \cite{Shenar+2017}. 
They argue that non-homogeneous evolution would have led to Roche-lobe overflow and mass transfer that leads to mass ratios inconsistent with the $q\sim 1$ they infer. They conclude CHE induced by birth spin best describes the system.

\paragraph{R\,144} is a WR binary in the LMC very similar to R145, but with a shorter orbital period ($P_\mathrm{orb} \approx 74$ days) and less extreme eccentricity ($e\approx0.5$); it is also more massive, having component masses of $74\pm4$\Msun and $69\pm4$\Msun \citep{Shenar+2021}. \cite{Shenar+2021} compare the components to evolutionary tracks from \citet{Brott+2011} and \citet{Kohler+2015}, and it is found that high initial rotation rates (birth-spin CHE) well-reproduce the observables except for the present day rotational velocities.

\paragraph{WR20a} is a very massive Milky Way WR binary, with component masses $\sim80$\Msun (with primary $82.7\pm5.5$\Msun and secondary $81.9\pm5.5$\Msun) and an orbital period of $3.686 \pm 0.01$ days \citep[][]{Bonanos+2004a, Rauw+2005,Bonanos+2010}. \citet{Rauw+2005} note that the spectrum of the system shows enhanced nitrogen and depleted carbon abundances at the stars' surfaces. They also estimate the velocity of rotation to be $\sim256$\,km\,s$^{-1}$ from the current orbital period (assuming synchronous rotation). This system has also been considered as a candidate for CHE evolution \citep{de-Mink+2008, de-Mink+2009}.

\paragraph{M33 X-7} is a massive O-star + BH binary \citep[with O star $70.0\pm6.9$\Msun and black hole $15.65\pm1.45$\Msun][]{Orosz+2007}, on an orbital period of 3.45 days, yet the O star remains within its Roche-lobe.  \citet{de-Mink+2009a} suggested CHE as an explanation, as this would allow the system to avoid Roche-lobe overflow (at least until the end of core helium burning). Alternatively, as proposed by \citet{Valsecchi+2010a}, this system may result from a phase of Roche lobe overflow where the main-sequence primary transfers part of its envelope to the secondary and loses the rest in a wind.

\section{Comparison of evolutionary tracks among the different input physics variations \label{app:comparison_tracks} }

\begin{figure}
\centering
\includegraphics[width=0.47\textwidth]{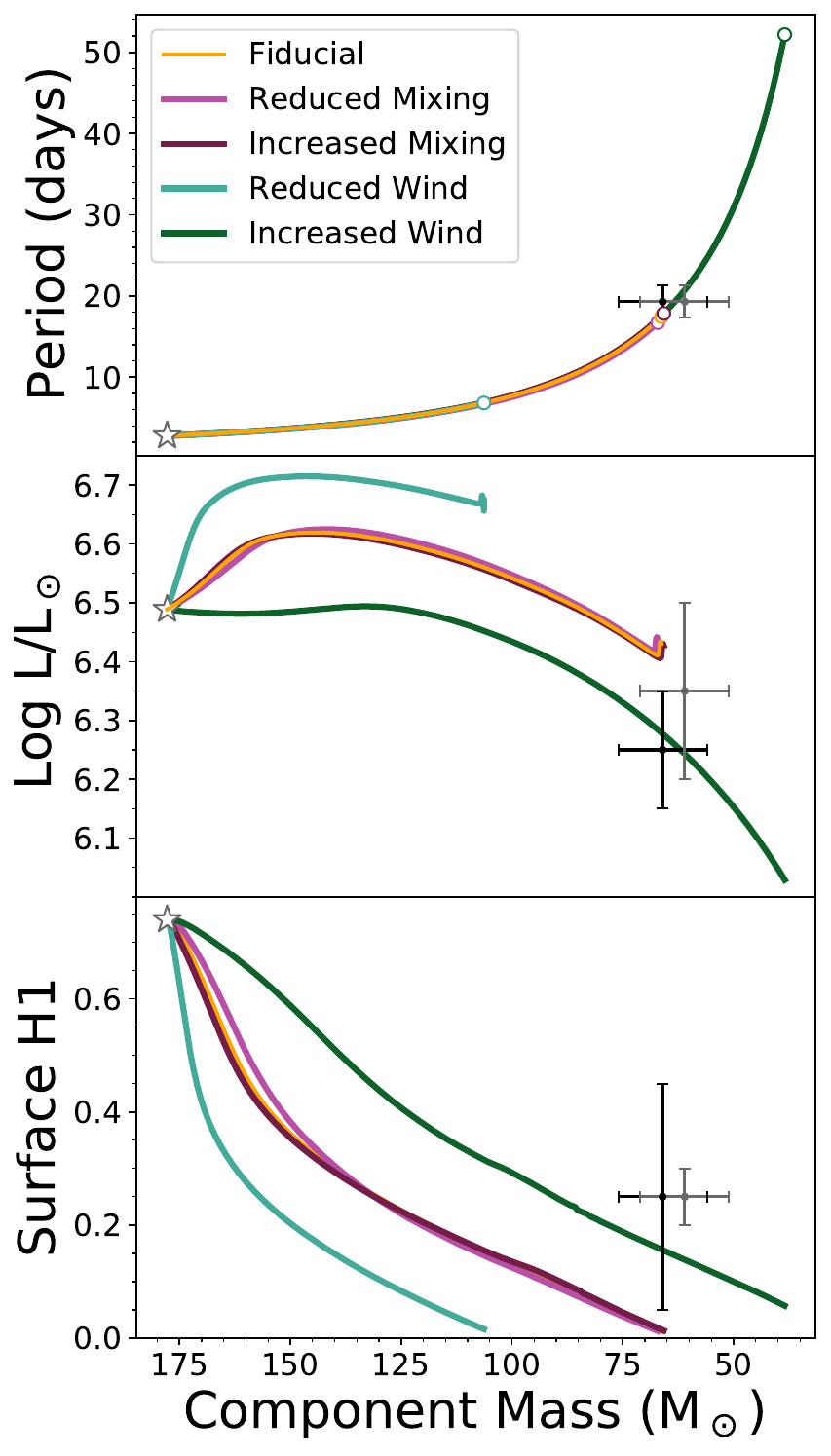}
\caption{Evolutionary tracks for all variations of the input physics, using models with initial mass $180\Msun$ and period 2.8 days. The fiducial model is shown as a solid orange line. The white star marks the beginning of the tracks; the stars evolve towards the right. Reduced and increased mixing are shown in pink and dark purple lines, respectively. Reduced and increased wind mass-loss are shown in teal and dark green. In the upper panel, a white circle shows the end point of the evolutionary track. Error bars show the observational values for HD\,5980's primary (gray) and secondary (black).}
\label{fig:phys_variations_evo}
\end{figure}

In Figure~\ref{fig:phys_variations_evo} we provide example evolutionary tracks for all five input physics variations for more insight into their evolutionary differences. To allow for a direct comparison, we fix the initial parameters to initial component masses of $180\Msun$ and an initial orbital period of 2.8 days. Note that the tracks shown here are thus not the best fitting models of their respective input physics variations grids. However, for this choice we are able to compare chemically homogeneously evolving models for all input physics variations. 

As can be seen in the upper panel, all tracks follow the same relation for the orbital widening as the system loses mass. However, the final orbital periods differ. The model with reduced mass loss widens fails to widen enough to explain the present-day orbital period of HD\,5980. This also explains the lack of good fits in the grid with reduced winds, see also Fig~\ref{fig:phys_variations_2Dpmd}. The three variations with fiducial mass-loss (decreased mixing, fiducial, and increased mixing) all widen to approximately HD\,5980's present day orbital period just before they terminate (at the beginning of core He burning). The increased mass-loss model widens far beyond the observed period.

In the middle panel we see that the evolution of the stars' luminosities. The models with reduced/enhanced mass loss are significantly brighter/dimmer than the fiducial, respectively. The models with different mixing variations are very similar to the fiducial model.

In the bottom panel we see the evolution of the surface hydrogen abundance. The models with increased/reduced mass loss show a shallower/steeper decrease of the surface hydrogen abundance as the star loses mass, respectively. This can be understood as being the results of two mechanisms that are responsible for enriching the surface.
The first mechanism is stripping by mass loss, which exposes deeper, chemically processed layers of the star. This process dominates for the model with increased mass-loss. The second mechanism is rotational mixing. This process dominates for the reduced mass loss model, which can be seen from the significant reduction of the hydrogen surface abundance down to 0.4 while losing only $10\Msun$. 

\begin{figure}
\centering
\includegraphics[width=0.99\linewidth]{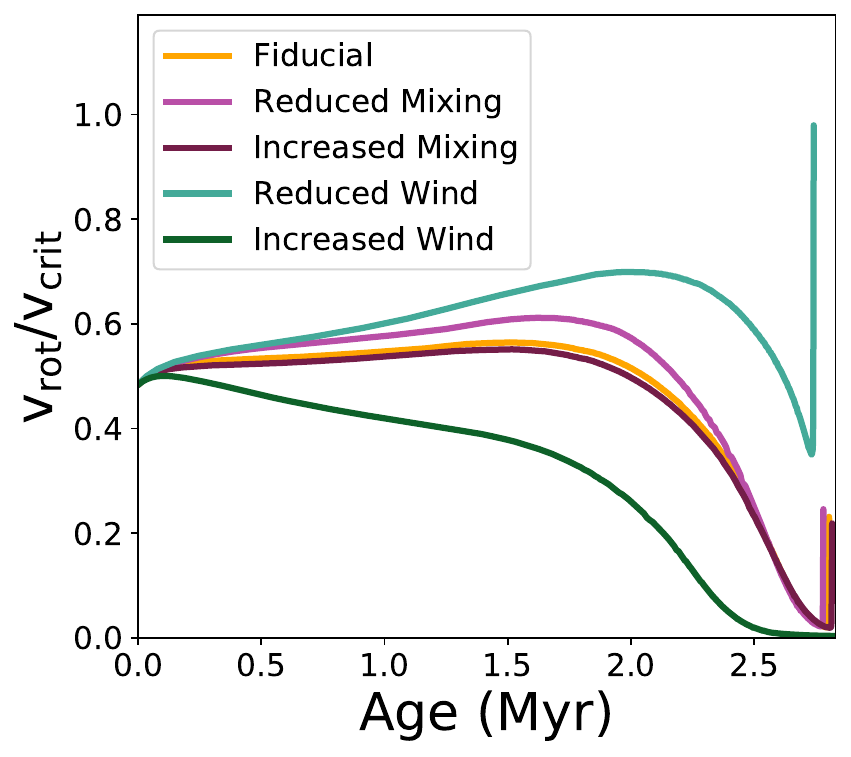}
\caption{The evolution of the rotation rate, expressed as a fraction of the critical velocity, as a function of age for each of the example models shown in Fig.~\ref{fig:phys_variations_evo}.}
\label{fig:vrot_vcrit}
\end{figure}

The reason that mixing is more efficient in the reduced mass loss model is due to its rotation --- which is a consequence of the difference in orbital evolution. Reduced mass loss reduces the widening, so tides force the star to co-rotate with a shorter orbital period. The stars thus spend a larger fraction of their lifetime spinning at a significant fraction of their critical velocity (see  Figure~\ref{fig:vrot_vcrit}). 

\section{Posterior distributions for the different input physics grids \label{app:posteriors var} }

\begin{figure}[ht!]
\centering
\includegraphics[width=0.98\linewidth]{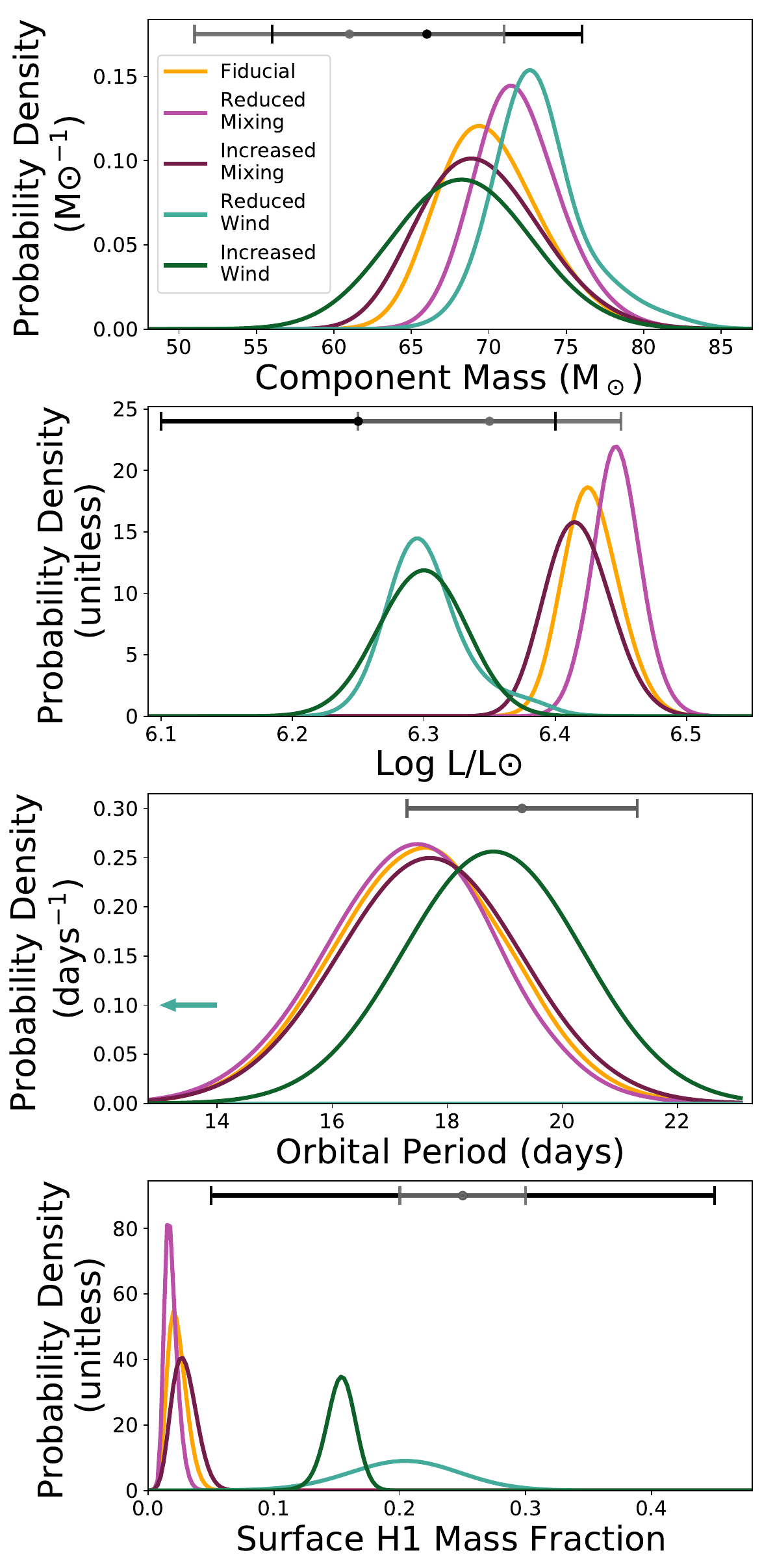}
\caption{Posterior probability distributions for all variations of the input physics on the observed parameters of stellar mass, luminosity, orbital period, and surface hydrogen mass fraction. Variations of the input physics are colored as in Figure~\ref{fig:phys_variations_evo}, and error bars representing the observed values for HD\,5980 A and B are shown at the top of each panel. Note that we use an error for the orbital period which is different from the observed standard deviation (see Section \ref{sec:bayesian_fit}). Additionally, in the panel for orbital period, the peak of the reduced wind distribution falls far below the lower bound of the plot --- it is represented by an arrow pointing leftward.} 
\label{fig:hist_obs}
\end{figure}

In Figure \ref{fig:hist_obs}, we show marginalized posterior probability distributions for all five variations of the input physics on each of the observed parameters. For most of these parameters, the peaks of the distributions for all input physics variations fall within 1-$\sigma$ of the observed values; all parameters, with the exception of surface H1 mass fraction, can be explained within 2-$\sigma$. Only the reduced and enhanced wind mass-loss variations are consistent with the surface abundances of H1 in HD\,5980 A and B to within 1- or 2-$\sigma$.

Comparing among the input physics variations, the enhanced mass loss distribution is always the best or second-best fit to the observed parameters, and thus the best overall. As denoted by the Bayes' factors in Section \ref{sec: bayes_factor}, enhanced wind mass-loss (and, thus, a combination of tidal CHE and wind stripping) is preferred in explaining the present-day state of HD\,5980\,A\&B.

\begin{figure}
\centering
\includegraphics[width=0.98\linewidth]{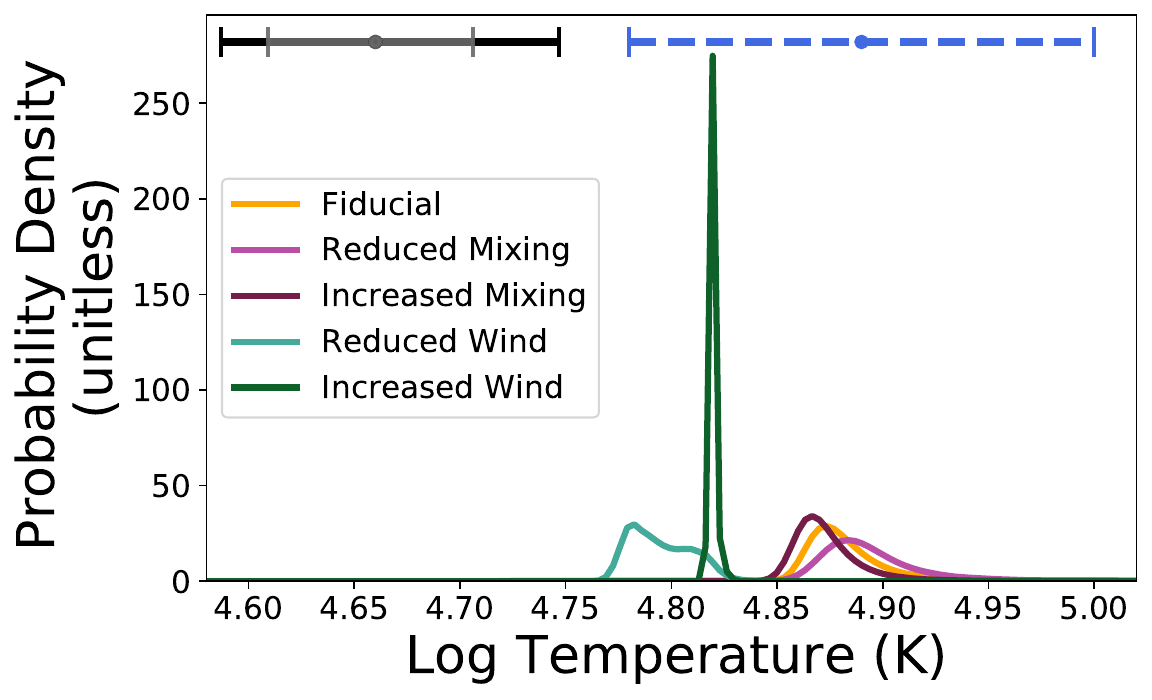}
\caption{Posterior probability distributions for the temperature. See the caption of Figure \ref{fig:hist_obs} for comments on coloration and error bars. Note that the model temperatures fall above the observed $T_\mathrm{eff}$ (the black and gray error bars), but are consistent with the value of the sonic-point temperature for star A (the blue, dashed error bar), log T$\approx4.8-5.0$ \citep[][Fig 16]{Georgiev+2011}.} 
\label{fig:hist_T}
\end{figure}

In Figure \ref{fig:hist_T}, we show the marginalized posterior probability distributions for effective temperature (on all five variations of the input physics). We include the observed surface temperatures for both stars quoted in \citet{Shenar+2016} as error bars at the top of the plot. The models' effective temperatures do not fit well to the reported observed surface temperature --- the distributions in Figure \ref{fig:hist_T} all fall above the reported temperature of $\sim45$\,kK. However, these temperatures are consistent with the temperature at the sonic point, $T_s$, from \cite{Georgiev+2011} (Figure 16), shown as a blue, dashed error bar at the top of the plot. As discussed in Section \ref{sec:limitations}, we believe $T_s$ is much closer to the temperature reported by \texttt{MESA}; it is encouraging that $T_s$ is consistent with our models even without fitting for it in our analysis.

\clearpage

\newpage
\bibliography{my_bib, my_bib2}
\bibliographystyle{aasjournal}

\end{document}